\documentclass[twocolumn,iop,numberedappendix,appendixfloats]{openjournal}
\usepackage[utf8]{inputenc}
\usepackage{amsmath,amssymb,amstext}
\usepackage{graphicx}
\graphicspath{ {./Figures/} }
\usepackage{bm}
\usepackage{natbib}
\usepackage{subcaption}
\usepackage[colorlinks,allcolors=blue]{hyperref}
\usepackage{dsfont}
\usepackage{comment}
\usepackage{float}
\usepackage{xcolor}
\usepackage{listings}
\usepackage{fontawesome}
\definecolor{maroon}{cmyk}{0, 0.87, 0.68, 0.32}
\definecolor{halfgray}{gray}{0.55}
\definecolor{ipython_frame}{RGB}{207, 207, 207}
\definecolor{ipython_bg}{RGB}{247, 247, 247}
\definecolor{ipython_red}{RGB}{186, 33, 33}
\definecolor{ipython_green}{RGB}{0, 128, 0}
\definecolor{ipython_cyan}{RGB}{64, 128, 128}
\definecolor{ipython_purple}{RGB}{170, 34, 255}

\lstdefinelanguage{iPython}{
    morekeywords={access,and,break,class,continue,def,del,elif,else,except,exec,finally,for,from,global,if,import,in,is,lambda,not,or,pass,print,raise,return,try,while},%
    morekeywords=[2]{abs,all,any,basestring,bin,bool,bytearray,callable,chr,classmethod,cmp,compile,complex,delattr,dict,dir,divmod,enumerate,eval,execfile,file,filter,float,format,frozenset,getattr,globals,hasattr,hash,help,hex,id,input,int,isinstance,issubclass,iter,len,list,locals,long,map,max,memoryview,min,next,object,oct,open,ord,pow,property,range,raw_input,reduce,reload,repr,reversed,round,set,setattr,slice,sorted,staticmethod,str,sum,super,tuple,type,unichr,unicode,vars,xrange,zip,apply,buffer,coerce,intern, function, @model, Uniform, Normal, MvNormal, theory_planck},%
    sensitive=true,%
    morecomment=[l]\#,%
    morestring=[b]',%
    morestring=[b]",%
    morestring=[s]{'''}{'''},% used for documentation text (mulitiline strings)
    morestring=[s]{"""}{"""},% added by Philipp Matthias Hahn
    morestring=[s]{r'}{'},% `raw' strings
    morestring=[s]{r"}{"},%
    morestring=[s]{r'''}{'''},%
    morestring=[s]{r"""}{"""},%
    morestring=[s]{u'}{'},% unicode strings
    morestring=[s]{u"}{"},%
    morestring=[s]{u'''}{'''},%
    morestring=[s]{u"""}{"""},%
    literate=
    {á}{{\'a}}1 {é}{{\'e}}1 {í}{{\'i}}1 {ó}{{\'o}}1 {ú}{{\'u}}1
    {Á}{{\'A}}1 {É}{{\'E}}1 {Í}{{\'I}}1 {Ó}{{\'O}}1 {Ú}{{\'U}}1
    {à}{{\`a}}1 {è}{{\`e}}1 {ì}{{\`i}}1 {ò}{{\`o}}1 {ù}{{\`u}}1
    {À}{{\`A}}1 {È}{{\'E}}1 {Ì}{{\`I}}1 {Ò}{{\`O}}1 {Ù}{{\`U}}1
    {ä}{{\"a}}1 {ë}{{\"e}}1 {ï}{{\"i}}1 {ö}{{\"o}}1 {ü}{{\"u}}1
    {Ä}{{\"A}}1 {Ë}{{\"E}}1 {Ï}{{\"I}}1 {Ö}{{\"O}}1 {Ü}{{\"U}}1
    {â}{{\^a}}1 {ê}{{\^e}}1 {î}{{\^i}}1 {ô}{{\^o}}1 {û}{{\^u}}1
    {Â}{{\^A}}1 {Ê}{{\^E}}1 {Î}{{\^I}}1 {Ô}{{\^O}}1 {Û}{{\^U}}1
    {œ}{{\oe}}1 {Œ}{{\OE}}1 {æ}{{\ae}}1 {Æ}{{\AE}}1 {ß}{{\ss}}1
    {ç}{{\c c}}1 {Ç}{{\c C}}1 {ø}{{\o}}1 {å}{{\r a}}1 {Å}{{\r A}}1
    {€}{{\EUR}}1 {£}{{\pounds}}1
    {^}{{{\color{ipython_purple}\^{}}}}1
    {=}{{{\color{ipython_purple}=}}}1
    {+}{{{\color{ipython_purple}+}}}1
    {-}{{{\color{ipython_purple}-}}}1
    {*}{{{\color{ipython_purple}$^\ast$}}}1
    {/}{{{\color{ipython_purple}/}}}1
    {+=}{{{+=}}}1
    {-=}{{{-=}}}1
    {*=}{{{$^\ast$=}}}1
    {/=}{{{/=}}}1,
    literate=
    *{-}{{{\color{ipython_purple}-}}}1
     {?}{{{\color{ipython_purple}?}}}1,
    identifierstyle=\color{black}\ttfamily,
    commentstyle=\color{ipython_cyan}\ttfamily,
    stringstyle=\color{ipython_red}\ttfamily,
    keepspaces=true,
    showspaces=false,
    showstringspaces=false,
    rulecolor=\color{ipython_frame},
    frameround={t}{t}{t}{t},
    numbers=none,
    numberstyle=\tiny\color{halfgray},
    backgroundcolor=\color{ipython_bg},
    basicstyle=\ttfamily\footnotesize,
    columns=fullflexible,
    keywordstyle=\color{ipython_green}\ttfamily,
}

\def\Planck{\textit{Planck}}

\usepackage{amssymb}
\usepackage{comment}

\newcommand{\vnh}{\hat{\mathbf{n}}}
\newcommand{\nobs}{n_g^{\rm obs}}
\newcommand{\nbar}{\bar{n}_g}
\newcommand{\valpha}{\mbox{\boldmath$\alpha$}}
\newcommand{\Mbar}{\bar{M}}
\newcommand{\vMbar}{\mbox{\boldmath$\bar{M}$}}
\newcommand{\vM}{\mbox{\boldmath$M$}}  
\newcommand{\vzeta}{\mbox{\boldmath$\zeta$}}  
\newcommand{\veta}{\mbox{\boldmath$\eta$}} 
\newcommand{\vbeta}{\mbox{\boldmath$\beta$}}   
\newcommand{\vMsys}{\mbox{\boldmath$M$}}   
\newcommand{\vepsilon}{\mbox{\boldmath$\epsilon$}}

\usepackage{graphicx}	% Including figure files
\usepackage{amsmath}	% Advanced maths commands
 \usepackage{amssymb}	% Extra maths symbols
\usepackage{bm}  % for bold math symbols

\def\aj{{AJ}}                   % Astronomical Journal
\def\apj{{ApJ}}                 % Astrophysical Journal
\def\apjl{{ApJ}}                % Astrophysical Journal, Letters
\def\apjs{{ApJS}}               % Astrophysical Journal, Supplement
\def\ao{{Appl.~Opt.}}           % Applied Optics
\def\aap{{A\&A}}                % Astronomy and Astrophysics
\def\jcap{{J. Cosmology Astropart. Phys.}}
\def\mnras{MNRAS}             % Monthly Notices of the RAS
\def\prd{{Phys.~Rev.~D}}        % Physical Review D
\def\pasj{{PASJ}}               % Publications of the ASJ
\date{\today}

\begin{document}
\journalinfo{The Open Journal of Astrophysics}
\submitted{submitted December 2024; accepted XXXX}

\shorttitle{J-PLUS. Additive versus Mutiplicative systematics in LSS surveys}
\shortauthors{Hernández-Monteagudo \& J-PLUS collaboration}
\title{The J-PLUS collaboration. Additive versus multiplicative systematics in surveys of the large scale structure of the Universe.}
%\subtitle{This is a SUBTITLE}

%\titlerunning{J-PLUS. Additive versus Mutiplicative systematics in LSS surveys}
% \authorrunning{C.Hernández-Monteagudo et al.}
   %\title{}
   %\subtitle{I. Overviewing the $\kappa$-mechanism}

\author{C.~Hern\'andez-Monteagudo$^{\star 1,2}$,
 G.~Aricò$^{19}$,
 J.~Chaves-Montero$^{18}$,
 L.~R.~Abramo$^{5}$,
 P.~Arnalte-Mur$^{13,14}$,
 A.~Hernán-Caballero$^{3,4}$,
 F.~J.~Galindo-Guil$^{3}$,
 C.~López-Sanjuan$^{3,4}$,
 V.~Marra$^{6,7,8}$,
 R.~von~Marttens$^{9}$,
 E.~Tempel$^{12}$,
 J.~Cenarro$^{3,4}$, 
 D.~Cristóbal-Hornillos$^{3}$,
 A.~Marín-Franch$^{3,4}$,
 M.~Moles$^{3}$,
 J.~Varela$^{3}$,
 H.~Vázquez Ramió$^{3,4}$,
 J.~Alcaniz$^{11}$,
 R.~Dupke$^{11}$,
 A.~Ederoclite$^{3,4}$,
 L.~Sodré~Jr.$^{15}$, 
\and R.~E.~Angulo$^{16,17}$}
\thanks{$^\star$ E-mail: \nolinkurl{chm@iac.es} }

\affiliation{$^1$ Instituto de Astrofísica de Canarias (IAC), 
              C/ Vía Láctea, S/N, E-38205, San Cristóbal de La Laguna, Tenerife, Spain} 
\affiliation{$^2$ Departamento de Astrofísica, Universidad de La Laguna (ULL), Avenida Francisco Sánchez,
             E-38206, San Cristóbal de La Laguna, Tenerife, Spain}
\affiliation{$^3$ Centro de Estudios de Física del Cosmos de Aragón,
            Plaza San Juan, 1, planta 3, E-44001, Teruel, Spain}
\affiliation{$^4$ Unidad Asociada CEFCA-IAA, CEFCA, Unidad Asociada al CSIC por el IAA, 
            Plaza San Juan 1, 44001 Teruel, Spain.}
\affiliation{$^5$ Departamento de Física Matemática, Instituto de Física, Universidade de São Paulo, R. do Matão 1371, 05508-090, São Paulo, SP, Brazil}
\affiliation{$^6$  Departamento de Física, Universidade Federal do Espírito Santo, 29075-910, Vitória, ES, Brazil}
\affiliation{$^7$ INAF -- Osservatorio Astronomico di Trieste, via Tiepolo 11, 34131 Trieste, Italy}
\affiliation{$^8$ IFPU -- Institute for Fundamental Physics of the Universe, via Beirut 2, 34151, Trieste, Italy}
\affiliation{$^9$ Instituto de Física, Universidade Federal da Bahia, 40210-340, Salvador-BA, Brazil}
\affiliation{$^{10}$ PPGCosmo, Universidade Federal do Espírito Santo, 29075-910, Vitória, ES, Brazil}
\affiliation{$^{11}$ Observatório Nacional, Rua General José Cristino 77, Rio de Janeiro, RJ, 20921-400, Brazil}
\affiliation{$^{12}$ Tartu Observatory, University of Tartu, Observatooriumi 1, Tõravere 61602, Estonia}
\affiliation{$^{13}$ Observatori Astronòmic de la Universitat de València, Ed. Instituts d'Investigació, Parc Científic, C/ Catedrático José Beltrán, n2, 46980, Paterna, Valencia, Spain}
\affiliation{$^{14}$ Departament d'Astronomia i Astrofísica, Universitat de València, 46100, Burjassot, Spain}
\affiliation{$^{15}$ Instituto de Astronomia, Geofísica e Ciências Atmosféricas, Universidade de São Paulo, 05508-090, São Paulo, Brazil}
\affiliation{$^{16}$ Donostia International Physics Centre (DIPC), Paseo Manuel de Lardizabal 4, 20018, Donostia-San Sebastián, Spain}
\affiliation{$^{17}$ IKERBASQUE, Basque Foundation for Science, 48013, Bilbao, Spain}
\affiliation{$^{18}$  Institut de Física d'Altes Energies, The Barcelona Institute of Science and Technology, Campus UAB, E-08193, Bellaterra, Barcelona, Spain}
\affiliation{$^{19}$ Institut für Astrophysik (DAP), Universität Zürich, Winterthurerstrasse 190, 8057, Zurich, Switzerland\\}

\begin{abstract}
Observational and/or astrophysical systematics modulating the observed number of luminous tracers can constitute a major limitation in the cosmological exploitation of surveys of the large scale structure of the universe. Part of this limitation arises on top of our ignorance on how such systematics actually impact the observed galaxy/quasar fields.
In this work we develop a generic, hybrid model for an arbitrary number of systematics that may modulate observations in both an additive and a multiplicative way, after applying a nonlinear power law transformation. This model allows us devising a novel algorithm that addresses the identification and correction for either additive and/or multiplicative contaminants. We test this model on galaxy mocks and systematics templates inspired from data of the third data release of the {\it Javalambre Photometric Local Universe Survey} (J-PLUS). We find that our method clearly outperforms standard methods that assume either an additive or multiplicative character for all contaminants in scenarios where both characters are actually acting on the observed data. In simpler scenarios where only an additive or multiplicative imprint on observations is considered, our hybrid method does not lie far behind the corresponding simplified, additive/multiplicative methods. Nonetheless, in scenarios of mild/low impact of systematics, we find that our hybrid approach converges towards the standard method that assumes additive contamination, as predicted by our model describing systematics. Our methodology also allows for the estimation of biases induced by systematics residuals on different angular scales and under different observational configurations, although these predictions necessarily restrict to the subset of {\em known/identified} potential systematics, and say nothing about ``unknown unknowns" possibly impacting the data.
\end{abstract}

%\maketitle

\section{Introduction}
\label{sec:intro}

Over the last four decades, surveys of the Large Scale Structure (LSS) of the Universe have dramatically improved in depth, area, and amount of information extracted from the sky. While first spectroscopic surveys like the CfA survey \citep{CfAsurvey} or {\it Las Campanas} \citep{LasCampanasSurvey} covered a moderate sky fraction of few hundred square degrees, yielding the position and distance of ${\cal O}[10^3-10^4]$ objects, the photometric surveys at that time, either in the optical \citep[e.g., APM,][]{APM_survey}, infrared \citep[e.g., 2MASS,][]{2mass_survey}, or radio (e.g., FIRST, \citet{first_survey}, NVSS \citet{nvss_survey}) would provide catalogues with ${\cal O}[10^6]$ objects in one or few frequency bands. This situation changed notably at the turn of the century with automatized surveys like SDSS, 2dF, or 6dF, \citep{sdss_survey, 2dF_survey, 6dF_survey}, which managed to cover thousands of square degrees, catalogue ${\cal O}[10^6-10^7]$ galactic and extra-galactic sources, and provide ${\cal O}[10^5-10^6]$ spectra. 

Currently ongoing and upcoming LSS surveys keep pushing frontiers further, again in terms of both depth and quality of the information obtained from each source. These LSS surveys can largely be split into two different categories: {\it (i)}
 photometric surveys like DES \citep{DES_survey}, HSC \citep{HSC_survey}, or Euclid \citep{euclid_survey}, typically conducted from large telescopes that enable state-of-the-art photometric depths, and with excellent image quality that allow mining into the deep and low surface brigthness universe while measuring shapes of galaxies in exquisite detail; and {\it (ii)} spectroscopic surveys, which attempt measuring high-quality spectra of $10^6$--$10^7$ of objects, like, e.g., HETDEX \citep{hetdex_survey}, Euclid\footnote{Euclid is a space mission conducting from Lagrange point L2 both extremely clean photometry and infra-red spectroscopy, see \url{https://www.euclid-ec.org/?page_id=2581}.} \citep{euclid_survey}, DESI \citep{desi_survey}, or 4MOST \citep{dejong12,4most_survey}. 

 In the recent years an intermediate type of surveys has appeared, in which both photometric depth and indiscriminate spectral information of all sources are targeted. This new type of surveys, dubbed as {\em spectro-photometric} surveys, typically consists on a set of relatively high (20--50) number of narrow or medium-width pass-band filters, each of which covers the entire footprint of the survey. These surveys also include some (deep) broad-band detection band, like the $r$ or $i$ optical bands. This results in moderately deep catalogues of sources with {\it pseudo-spectra} of resolution factor $R\sim 20$--$50$, depending on the number of pass-band filters used. Although having to observe in multiple bands limits to some extent their photometric depth, the lack of target pre-selection and the wealth of information contained in those pseudo-spectra make this type of surveys unique for the large range of interest of their data. While surveys like COMBO-17 \citep{combo17_survey}, ALHAMBRA \citep{alhambra_survey,alhambra_survey_II}, COSMOS \citep{cosmos_survey}, MUSYC \citep{musyc_survey}, CLASH \citep{clash_survey}, or SHARDS \citep{shards_survey} are regarded as the pathfinders, and  more recent efforts like, PAU \citep{pau_survey}, J-PLUS \citep{jplusIntropaper}, J-PAS \citep{BenitezJPAS,miniJPAS_survey}, and even SPHEREx \citep{sphereX_survey} constitute the state-of-the-art of this observational strategy. 

The motivation behind mapping the LSS includes not only astro-physical reasons, but also invokes key questions in Cosmology and Fundamental Physics of paramount importance, such as what is the nature of ``dark" fundamental components of the universe like dark matter and dark energy, what is the number and the mass of relativistic species in nature, or what is the physics in the initial inflationary epoch that gave rise to the structure we see in the universe today. LSS surveys also allow testing gravity against Einstein's General Relativity on cosmological scales and constrain extensions of this theory. Furthermore they can also shed light on open astrophysical questions such as how galaxies (and the stellar populations therein) were formed and evolved, what is the mass and age distribution of black holes in the universe, and what generated their seeds so that they were present in such early stages of cosmological history. 

The physical interpretation of most (if not all) surveys of the LSS are limited, at some degree, by some level of known or unknown systematics. For instance, in surveys like SDSS/BOSS, the mis-identification of galaxies with stars in photometric surveys, and the modulation of the observed galaxy field by the light due to unresolved star emission from our Galaxy have been identified as some of the main culprits for our limitations when interpreting the largest scales, \citep{boss_syst1,boss_syst2,boss_syst3,boss_chm}. Initially, while studying the DES Science Verification data, \citet{Leistedt_DESSV16} identified as seeing variations the main systematics source in clustering and lensing analyses, concluding it could be kept under control. However, to date, there is yet no DES data analysis interpreting the survey on its largest scales (despite covering about $5,000$~sq.deg.), and in the literature there are relatively few works thoroughly addressing the study on clustering on the very large distances, and in practically all those cases the presence of systematics is the main limitation. 

There is a wealth of methods attempting to correct galaxy and quasar catalogues from systematics in LSS surveys \citep[][to quote just a few]{boss_syst1,boss_syst2,boss_syst3,boss_chm,leistedt_and_peiris_2013,leistedt_and_peiris_2014,Leistedt_DESSV16,delubac17,ross2017,elvin-poole18,xavier19,kong_eboss20,Wagoner_DESy1,rmonroy_desy3}. They make use of knowledge of observing conditions like average seeing in each footprint pixel, airmass, sky background, etc, together with other potential systematics of astro-physical origin, like Milky Way stellar density, galactic extinction, zodiacal light, etc. All these measurables are compared to the observed galaxy/qso density map, and from this comparison some type of correction is applied. In the insightful and thorough work of \citet{weave+21}, it has been shown that most of the methods in the literature can be related to an Ordinary Linear Squared (OLS) exercise by which 
a signal of a given form (the modulation due to systematics) is extracted out of a background of Gaussian {\it noise} (which corresponds to the desired, ``true", galaxy/quasar field). On top of other assumptions which may or may not completely hold (Gaussian and un-correlated character of the galaxy/quasar field), the OLS also assumes that the impact of systematics is some type of spurious signal that is {\em added} on top of the real one. \citet{weave+21} modify the OLS to account for multiplicative systematics (i.e., systematics that impact the real, underlying galaxy/quasar field in a multiplicative way), and they actually assume this multiplicative character throughout their work. 

In our work we drop this assumption, and devise a new methodology for systematics amelioration that incorporates the OLS power without making any assumption on the character of systematics (whether they act in an additive or multiplicative manner). Our method does however rely on a model for systematics which, despite of attempting to be as generic and as realistic as possible, may or may not reflect precisely the real impact of systematics on a given survey. This work is structured as follows: in Sect.~\ref{sec:jplusdr3} we introduce the data of J-PLUS Data Release 3 (DR3), data that we use as reference for building out set of systematics templates and our log-normal mocks for galaxy density fields. In Sect.~\ref{sec:model} we describe our model of systematics, while in Sect~\ref{sec:method} we introduce our systematics-correction method. We further tests its performance in Sect.~\ref{sec:results}, first on J-PLUS DR3 motivated mocks, and then briefly on real J-PLUS DR3 data. We discuss and provide further insight of our results in Sect.~\ref{sec:discussions}, and summarize our conclusions in Sect.~\ref{sec:conclusions}.

\section{J-PLUS DR3}
\label{sec:jplusdr3}
%_____________________
\begin{figure*}
\hspace*{-4.cm}
\includegraphics[scale=0.55]{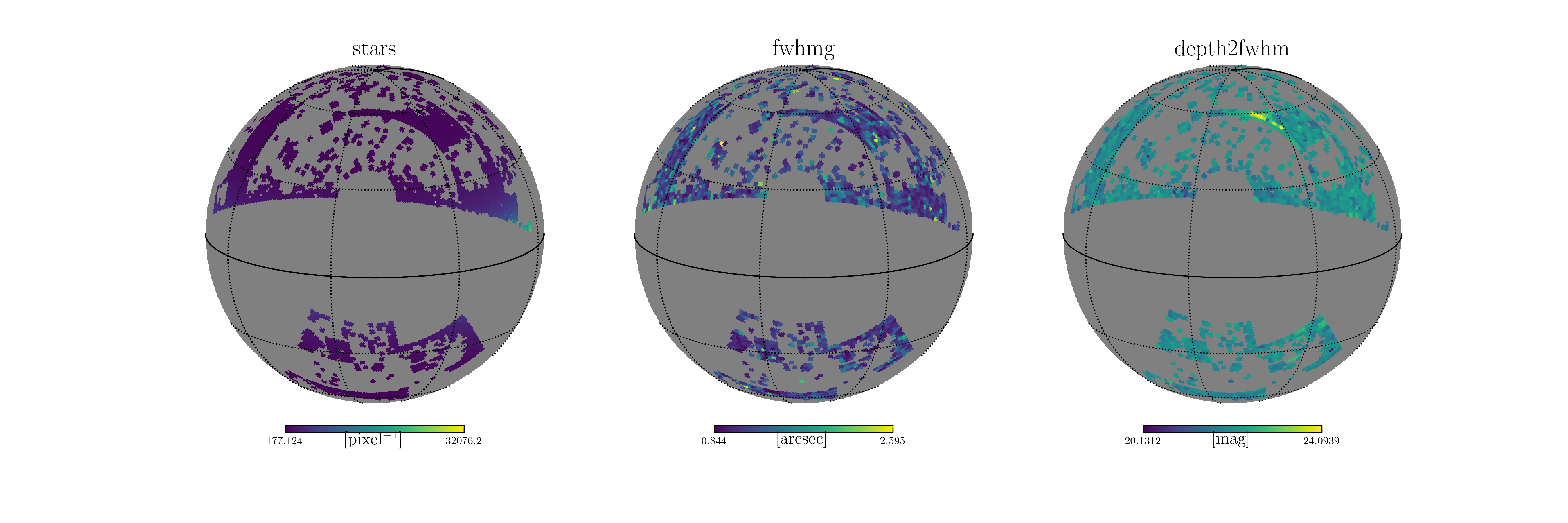}
\caption{Orthographic projections of  three different J-PLUS DR3 potential systematics, namely the star density (left panel, {\tt stars}), the {\tt SExtractor} FWHM estimate assuming a Gaussian shape of the point source (middle panel, {\tt fwhmg}), and the magnitude at SNR$=5$ and 2$\times$FWHM aperture (right panel, {\tt depth2fwhm}). }
\label{fig:map_syst1}
\end{figure*}
%-------------------------

We shall test our systematics correction algorithms on galaxy data and survey observing conditions motivated from the {\it Javalambre Photometric Local Universe Survey} \citep[hereafter J-PLUS,][]{jplusIntropaper}.
This survey covers the northern sky from an 83~cm diameter optical telescope at the {\it Observatorio Astrofísico de Javalambre} (OAJ\footnote{OAJ URL site: \url{https://oajweb.cefca.es/}}) since mid 2017. This Javalambre Auxiliary Survey Telescope (JAST/T80) has a German-equatorial mount with a field-of-view diameter of 2~deg, and mounts T80Cam, a camera with one 9.2~kpix$\times$9.2~kpix CCD designed by Teledyne e2V (UK) of readout times of $\sim 12$~s with typical read-out noise levels of $3.4$~e$^-$ (RMS). J-PLUS is unique for the optical system it mounts: on top of standard, Sloan-type $u,g,r,i,$ and $z$ broad-band filters, the optical system includes 7 other medium-width and narrow band filters (of $200$--$400$~\AA\, width) centered on particular spectral features like the Ca H+K lines (the $J0395$ filter), H$\delta$ ($J0410$), the G band ($J0430$), the Mg$b$ triplet ($J0515$), and the Ca triplet ($J0861$). The two additional $J0378$ and $J0660$ narrow filters are sensitive to the [OII]/$\lambda$3727 and H$\alpha$/$\lambda$6563 lines, respectively.  Given the size of the telescope, and the {\cal O}[$10^3$~s] duration exposures, the limit magnitudes reached for the $g,r,$ and $i$ broad bands are $22,22$, and $21.75$, respectively, while for the $u$ and $z$ bands those are $21$ and $20.75$, respectively. The depth in the narrower filters is not very different, with typical limit magnitudes at the level of $21$, becoming shallower for the redder filters ($20.5$ for $J0861$). We refer the reader to the work of \citet{jplusIntropaper} for further technical details on this survey. 
\\

While this filter system allows stellar population studies in the local universe, including our own galaxy, it also enables the identification of bright line emitters on particular lines whose wavelengths are redshifted on any of the narrow-band filters in the tray attached to T80Cam \citep{jplus_emitters_daniele,jplus_alejandro}. In general, one expects that this extra set of narrow/medium width optical filters to significantly improve the accuracy and precision of photometric redshifts (hereafter photo-$z$s), provided that {\em every} entry in the catalogue will have a $R\sim 10$ spectrum. We typically expect photo-$z$ errors at the level of $0.01$--$0.03$~\% \citep{jplusIntropaper}, but this will be addressed in more detail when formally studying the clustering of J-PLUS galaxies \citep{chm-jplus-tomoDR3}. For data release 3 (DR3), J-PLUS has covered about $3\,192$~sq.deg. ($2\,881$~sq.deg after masking), and contains about 47.4 million sources in the dual catalogue (i.e., sources being at least detected in the $r$-band, of which $\sim 29.8$~million have $r\leq 21$). \\

All the relevant photometric information (including derived quantities like photo-$z$ probability density functions) of  J-PLUS sources are archived at the site \url{https://archive.cefca.es/catalogues}.
This database also hosts information about observing conditions like air mass, exposure time, instrumental noise, Milky Way extinction, number of exposures in each tile, etc, which are needed to build angular templates of systematics potentially impacting the observed number of galaxies. A complete list of the 14 potential systematics one can extract from the database is given in Table~\ref{tab:syst1}. In orthographic projections it can be clearly seen that some of these templates are strongly correlated in their angular pattern on the celestial sphere, and this must be accounted for in our forthcoming analyses. For display purposes, in Fig.~\ref{fig:map_syst1} we show the angular pattern of three systematics templates used in this work in pixels of $\sim 27~$ arcmin size, namely the star density (left panel), the {\tt SExtractor} FWHM estimate after assuming that the point spread function (PSF) in the $r$ band is a Gaussian (middle panel), and the magnitude corresponding to SNR$=5$ under an aperture equal to twice the FWHM of the PSF (also in the $r$-band image, right panel). These three templates show seemingly independent angular patterns, a situation that not always holds among other templates. For instance, both extinction and star-density templates are highly correlated, and depth-related templates like {\tt noise, teffective, texposed} and {\tt ncombined} display certain similarities as well. It is thus evident that properly accounting for the angular correlation between potential systematics templates will be required when characterizing and minimizing their impact on the observed galaxy density field.\\

%A more detailed description of the potential systematics and their angular correlation can be found in Appendix~\ref{sec:appA}.
%Final labels=  ['stars', 'zpt', 'noise', 'efectime', 'texposed', 'ncombined', 'fwhmg', 'm50s', 'depth2fwhm', 'depth3as', 'ebv', 'ebvPlanck', 'airmass', 'odds']
\begin{table}
\caption{\label{tab:syst1} Full set of observables upon which we build our potential systematics templates. A brief description for some of the templates follows. Templates with indexes $1\rightarrow 9$ and $12$ have been built from information obtained from the $r$-band co-added images (with $r$-band acting as the detection band in J-PLUS). Index number 6 refers to the FWHM measured by {\tt SExtractor} under the approximation that the PSF is Gaussian, while index 7 provides the magnitude at which 50~\% completeness is reached for point sources. Indexes 8 and 9 indicate the magnitude reached at SNR=5 under different apertures (twice the FWHM and 3~arcsec, respectively). Indexes 10 and 11 refer to two very similar templates providing the E(B-V) colour excess from \citet{Schlegel1998} and the {\it Planck} mission\footnote{Visit {\tt https://pla.esac.esa.int/} for accessing its public data.}. Finally, index 13 refers to average {\it odds} parameter from all galaxies falling in each pixel. The {\it odds} is a proxy for the compactness of the probability density distribution of the photometric redshift, as it will be shown below. This latter template is likely to be correlated to any other template related to the photometric quality and depth. }
%\begin{ruledtabular}
\begin{tabular}{ccc}
Index & Observable & Label \\
\hline
0 & Star density & {\tt stars} \\
1 & Zero-point of the image & {\tt zpt} \\
2 & Estimated image noise & {\tt noise}\\
3 & Effective total exposure time  & {\tt teffective} \\
4 & Total exposure time  & {\tt texposed}\\
5 & Number of reduced images combined  & {\tt ncombined} \\
6 & FWHM estimate under Gaussian PSF   & {\tt fwhmg} \\ 
7 & 50\% detection mag. for point-like sources & {\tt m50s} \\
8 & Mag. at SNR=5 and 2$\times$FWHM aperture & {\tt depth2fwhm} \\
9 & Mag. at SNR=5 and 3~arcsec aperture & {\tt depth3as} \\
10 & E(B-V) colour excess from SFD98 & {\tt ebv} \\
11 & E(B-V) colour excess from {\Planck}  & {\tt ebvPlanck} \\
12 & Average airmass of tile & {\tt airmass} \\
13 & Galaxy-weighted average odds parameter & {\tt odds} \\
\hline
 \end{tabular}
%\end{ruledtabular}
\end{table}

In our analyses, we conveniently re-scale these templates in the following way. We fit a power law for the approximate scaling of the observed number of galaxies ($\nobs$) with respect to the $j$-th systematics map, i.e., we find the power law index $\alpha^j_s$ for the scaling
\begin{equation}
\frac{\nobs (\vnh)}{\langle \nobs \rangle_{\vnh}} \propto \biggl(\frac{M_j(\vnh)}{\langle M_j\rangle_{\vnh}}\biggr)^{\alpha^j_s}.
\label{eq:nobs_plscal}
\end{equation}

If $\alpha^j_s$ is found to be, in absolute value, greater than unity, then we re-scale this systematic map by 
\begin{equation}
M_j\rightarrow M_j^{1/|\alpha^j_s|}.    
\label{eq:Mrescl}
\end{equation}
 We otherwise leave $M_j(\vnh)$ untouched. In this way, we impose a smooth and moderate variation of the $j-$th systematics map wrt the observed angular number density, something convenient when expanding $M_j(\vnh)$ in terms of its angular anisotropies in harmonic space. 
 Also for convenience, we finally renormalize all systematic templates to have unity variance over the sky footprint, i.e., $M_j(\vnh)=(\bar{M}_j+ \delta M_j(\vnh))$, such that $\langle \delta M_j^2(\vnh) \rangle_{\vnh}=1$. 

%________________________________________________________________
\section{Theoretical modeling of systematics}
\label{sec:model}
In what follows, we try to model the observed angular number density of galaxies ($\nobs (\vnh)$) as a function of the true, underlying one ($\nbar(1+\delta_g)$), with $\delta_g (\vnh)$ the cosmological density contrast of a given galaxy population, and $\nbar$ its average number density. For this purpose, we  consider a family of systematics $M_i (\vnh)$, with $i=1,N_{s}$ and $N_s$ the total number of potential systematics that may bias the observed number density of galaxies. We choose to model the observed number density of galaxies as\footnote{This expression attempts to be a realistic model for the observed galaxy angular number density in terms of potential systematics, in which multiplicative systematics impact both the real, underlying galaxy field and the additive systematics. We believe this is more generic than considering additive systematics outside the influence of multiplicative ones.}
\begin{equation}
\nobs (\vnh) = \biggl( \nbar (1+\delta_g) + \valpha \cdot \vMsys \biggr) \prod_{1}^{N_s} (1+\beta_i \delta M_i),
\label{eq:smodel1}
\end{equation}
where the $i$-th systematic map is decomposed into its angular mean ($\Mbar_i$) and fluctuations ($\delta M_i$) as
\begin{equation}
M_i(\vnh) = \Mbar_i+ \delta M_i (\vnh),
\end{equation}
such that $\langle \delta M_i^2 (\vnh)\rangle_{\vnh}=1$ after a proper re-normalization. The components of the $N_s$-long vectors $\valpha$ and $\vbeta$ provide the amount of {\em additive} and {\em multiplicative} presence associated to the potential systematics map $M_i (\vnh)$ in $\nobs (\vnh)$. It is trivial to find that Eq.~\ref{eq:smodel1} can be rewritten as
\begin{equation}
\begin{split}
\nobs (\vnh)  = & \,\,  \nbar(1+\delta_g+\vepsilon\cdot\vMbar + \vbeta\cdot\veta) + \delta \vM \cdot \vzeta \,+   \\\
& \nbar\delta_g  (\vbeta \cdot \delta \vM) + \nbar \delta_g \sum_{i,j} \delta M_i \delta M_j \beta_i \eta_j \, + \\\
 & \sum_{i,j} \delta M_i \delta M_j \beta_i \eta_j  (1-\delta^{K}_{i,j}) + {\cal O}[3^{\rm rd} \,\mathrm{in}\,\vepsilon,\vbeta].
 \label{eq:smodel2} 
\end{split}
\end{equation}
By neglecting the cross terms involving $\delta_g$ with $\delta \vM$ (of order ${\cal O}[\delta_g \times \vMsys]$) we can simplify this equation further:
\begin{equation}
\begin{split}
\nobs (\vnh)  = & \,\, \nbar(1+\delta_g+\vepsilon\cdot\vMbar + \vbeta\cdot\veta) + \delta \vM \cdot \vzeta + \\\
 & \sum_{i,j} \delta M_i \delta M_j \beta_i \eta_j  (1-\delta^{K}_{i,j}) + {\cal O}[3^{\rm rd} \,\mathrm{in}\,\vepsilon,\vbeta] \\\
 & + {\cal O}[\delta_g \times \vMsys].
\label{eq:smodel3} 
\end{split}
\end{equation}
In both equations, we have introduced the vector $\vepsilon$ whose components read from $\alpha_i := \nbar \epsilon_i$, and account for the additive contribution to the observed galaxy number density. Likewise, the dot product involving the vectors $\delta \vM$ and $\vzeta$ can be rewritten as 
\begin{equation}
\delta \vM \cdot \vzeta = \sum_{i=1}^{N_s} \delta M_i \zeta_i,
\label{eq:dotMzeta}
\end{equation}
with $\zeta_i = \nbar (\epsilon_i + \beta_i [1+\vepsilon\cdot \vMbar])$, and $\Mbar_i$ the $i$-th component of $\vMbar$. At the same time, $\veta:=\nbar (\vepsilon + \vbeta)$. From these expressions it is easy to infer several statements: {\it (i)} an ordinary least square (OLS) regression will typically estimate the amplitude of $\vzeta$ but will not be able to distinguish between the additive or multiplicative nature of the systematics; {\it (ii)} additive systematics will bias the observed average number density of objects $\langle \nobs \rangle_{\vnh}$ {\em linearly} in $\vepsilon$ via the term $\nbar\,\vepsilon\cdot\bar{\vM}$, and multiplicative systematics will {\em also} bias $\langle\nobs \rangle_{\vnh}$ quadratically via the $\vbeta\cdot\veta$ term; and {\it (iii)} after removing an estimate for the linear term in $\delta \vM$ (i.e., $\vzeta\cdot\delta \vM$), the variance of the resulting galaxy field will be modulated by terms that are (at least) linear in the $\beta_i$'s (via e.g,. the cross term $\bar{n}_g \delta_g\,(\vbeta\cdot\delta \vM)$ in Eq.~\ref{eq:smodel2}), while additive systematics will have no impact in the resulting galaxy variance at linear order.   

One can expect that additive systematics are much better kept in control than multiplicative ones, since the source purity requirements in LSS surveys are usually very strict. However, estimating the purity of a given matter tracer sample may be far from trivial under several circumstances, like in regions of low galactic latitude where the star angular density becomes an issue. Provided the fact that the presence of additive systematics impacts linearly the estimation of the monopole/average of the galaxy number density, and that this bias impacts linearly the amplitude of the estimated galaxy density contrast field ($\delta_g^{\rm obs}= \nobs / \langle \nobs\rangle_{\vnh}-1$), we shall distinguish the different additive and multiplicative character of systematics in this work, in an effort to provide realistic bias and error forecasts. For this purpose, in the next section we design a {\em hybrid}, novel systematic correction/amelioration algorithm which attempts to solve for the additive ($\vepsilon$) and multiplicative ($\vbeta$) contribution of the systematics to the observed number density of galaxies.

\subsection{Handling of sky signals on the 2D sphere}

In this work we shall work in multipole (or Fourier) space of the 2D sphere. Any full sky signal $\mathbf{d}(\vnh)$ can be decomposed in harmonic space as
\begin{equation}
\mathbf{d} (\vnh) = \sum_{\ell=0}^{\ell_{\rm max}} d_{\ell,m} Y_{\ell,m} (\vnh),
\label{eq:mapdecomp1}
\end{equation}
where the $Y_{\ell,m} (\vnh)$ are the usual, scalar spherical harmonics and $d_{\ell,m}$ constitute the multipole coefficients of $\mathbf{d}(\vnh)$. One can write the product of two sky signals $\mathbf{d} (\vnh)$ and $\mathbf{m} (\vnh)$ as
\begin{eqnarray}
\nonumber
\langle \mathbf{d} \cdot \mathbf{m} \rangle_{\vnh} & = &\biggl \langle
\sum_{\ell_1,m_1} \sum_{\ell_2,m_2} d_{\ell_1,m_1} m^{\star}_{\ell_2,m_2} Y_{\ell_1,m_1}(\vnh) Y_{\ell_2,m_2}^{\star} (\vnh) \bigg \rangle_{\vnh} \\
 & = &\sum_{\ell_1} \frac{2\ell_1+1}{4\pi}\bigl \langle d_{\ell_1,m_1} m^{\star}_{\ell_1,m_1} \bigr \rangle_{m_1},
\label{eq:xcorr1}
\end{eqnarray}
where the $\star$ symbol denotes ``complex conjugate", the symbol $\langle (...) \rangle_{\vnh} = 1/(4\pi) \int d\vnh (...)$ denotes a normalized integral on the sphere, and the ensemble average $\langle (...) \rangle_{m} = 1/(2\ell+1) \sum_{m=-\ell}^\ell (...)$ expresses a sum over the magnetic or azymuthal number $m$ within the limits corresponding for a given multipole $\ell$. The ensemble average $\langle d_{\ell_1,m_1} m^{\star}_{\ell_1,m_1}\rangle_{m_1} $ is also known as the $\ell$-th multipole of the cross-angular power spectrum of $\mathbf{d}$ and $\mathbf{m}$, denoted by $C_{\ell}^{\mathbf{d},\mathbf{m}}$. Likewise, by replacing the multipoles of the $\mathbf{m}$ field by those of $\mathbf{d}$ one obtains the auto-power spectrum of the $\mathbf{d}$ field, $C_{\ell}^{\mathbf{d},\mathbf{d}}$.

Furthermore, if $\mathbf{n} (\vnh)$ is a noise field embedded in $\mathbf{d}$, one can compute a noise weighted product of the two sky signals $\mathbf{d}$ and $\mathbf{m}$ via
\begin{equation}
\langle \mathbf{d} \cdot \mathbf{m} \rangle_{\rm Nw} = \langle \mathbf{d} (\mathbf{N}(\vnh,\vnh))^{-1} \mathbf{m}\rangle_{\vnh},
\label{eq:weighted_prod}
\end{equation}
where $\mathbf{N} (\vnh,\vnh)$ is the ensemble average (through noise realisations) of the product of the noise fields at position $\vnh$, $\mathbf{N} (\vnh,\vnh) := \langle \mathbf{n}(\vnh)\,\mathbf{n} (\vnh) \rangle = \langle \mathbf{n}^2 (\vnh) \rangle$. Notice that, more generally, one can obtain a noise covariance matrix by looking at the noise field at two different sky positions, $\mathbf{N} (\vnh_1,\vnh_2) := \langle \mathbf{n}(\vnh_1)\,\mathbf{n} (\vnh_2) \rangle$. Likewise, one can generalise the product of Eq.~\ref{eq:weighted_prod} and write
\begin{equation}
\langle \mathbf{d} \cdot \mathbf{m} \rangle_{\rm Nw2} = \mathbf{d}^t \mathbf{N}^{-1} \mathbf{m},
\label{eq:weighted_prod2}
\end{equation}
where the sky signals are now vectors, with the subscript $t$ denoting ``transpose", and where all possible pairs of elements of $\mathbf{d}$ and $\mathbf{m}$ are weighted by the inverse noise covariance matrix $\mathbf{N}^{-1}$. The number of operations involved in Eq.~\ref{eq:weighted_prod2} is, a priori, much larger (${\cal O}[N_{\rm pix}^2]$) than those involved in Eq.~\ref{eq:weighted_prod} (${\cal O}[N_{\rm pix}]$), with $N_{\rm pix}$ the number of pixels or evaluations of the signals on the sky footprint). Note as well that the elements of the vectors $\mathbf{d}, \mathbf{m}$ can be either evaluations of these fields along different sky positions, $\mathbf{d}=\{ \mathbf{d}(\vnh_1), \mathbf{d}(\vnh_2), ...\}$, {\em or}, if we choose to work in Fourier/harmonic space, those elements can be multipole coefficients, $\mathbf{d}=\{ d_{\ell_1,m_1}, d_{\ell_1,m_2}, ..., d_{\ell_2,m_1},... \}$. Provided that operations in Fourier space are much more efficient thanks to fast Fourier transforms, in this work we shall adopt the latter, Fourier representation.  All these operations of signals on the 2D sphere are performed with the HEALPix~\footnote{HEALPix's URL site \url{http://healpix.sf.net}. Throughout this work we use the python package {\tt healpy}, \citet{healpy}.} \citep{healpix} software. We shall use the pixel resolution parameter $N_{\rm side}=128$, which corresponds to pixels of $\theta_{\rm pix}\sim 27$~arcmin on a side\footnote{For simplicity we stick to this value of $N_{\rm side}$, although our methodology has a priori no limitation in the value of $N_{\rm side}$ adopted. Higher values of $N_{\rm side}$ result in sensitivity to smaller angular scales, which are however impacted by non-linear physics whose modelling is not straightforward. } 

Finally, we shall assume that the cosmological galaxy density field $n_g(\vnh) $ is isotropic. This will significantly simplify our computations later on, since, as we shall show below, $n_g(\vnh) $ will be regarded as a noise field in the context of systematics removal. In Fourier space, isotropic signals have {\em diagonal} correlation matrices, i.e., $\langle n_{g,\ell_1,m_1} n^{\star}_{g,\ell_2,m_2}\rangle \propto \delta^K_{\ell_1,\ell_2} \delta^K_{m_1,m2}$, with $\delta^K_{i,j}$ the Kronecker delta, that is, $\delta^K_{i,j}=1$ for $i=j$ and $\delta^K_{i,j}=0$ otherwise ($i\neq j$).

%________________________________________________________________
\section{Methodology}
\label{sec:method}

\subsection{The standard additive and multiplicative approaches}

Most of approaches in the literature attempt to correct for systematics by adopting an ordinary least squares method \citep[e.g.,][ also known as a ``matched filter" approach in the context of studies of the Cosmic Microwave Background anisotropies]{weave+21,OliCosta1999}. Given a signal $\mathbf{m}$ whose profile is known, and which is embedded in a Gaussian noise field $\mathbf{n}$, conforming in a total, measured field $\mathbf{d}$ given by 
\begin{equation}
\mathbf{d} = \valpha\mathbf{m} + \mathbf{n},
\label{eq:ols1}
\end{equation}
it can be trivially shown that a minimization of the $\chi^2$ statistic
\begin{equation}
    \chi^2 = (\mathbf{d}-\valpha\mathbf{m})^t \mathbf{N}^{-1} (\mathbf{d}-\valpha\mathbf{m}) 
    \label{eq:chisq1}    
\end{equation}
with respect to the amplitude $\valpha$ yields the following, {\em minimum} variance estimates of $\alpha$ and its uncertainty:
\begin{equation}
    E[\valpha] = \frac{\mathbf{d}^t\,\mathbf{N}^{-1}\mathbf{m}}{\mathbf{m}^t\,\mathbf{N}^{-1}\mathbf{m}}; \phantom{xxxxxx} \sigma^2[\valpha]=(\mathbf{m}^t\,\mathbf{N}^{-1}\mathbf{m})^{-1}.
    \label{eq:ols2}
\end{equation}
In the equations above, the superscript $t$ denotes ``tranpose", and $\mathbf{N}$ refers to the covariance of the noise field $\mathbf{n}$. 

Despite the fact that the galaxy density field is not Gaussian (although it is not far from Gaussianity on the large scales), this methodology is directly exported to LSS surveys \citep[][]{leistedt_and_peiris_2013,leistedt_and_peiris_2014,Elsner_17,weave+21}: the observed, total signal $\mathbf{d}$ is identified with the observed galaxy density $n_g^{\rm obs}$, the Gaussian noise $\mathbf{n}$ with the real, underlying galaxy distribution $\bar{n}(1+\delta_g)$, and the known signal $\mathbf{m}$ with some potential systematic map given by $\mathbf{M}$. Even when the OLS is strictly built upon an additive contaminating signal (Eq.~\ref{eq:ols1}), for low enough contamination levels (or $\valpha$-amplitude values), one usually applies OLS for correcting multiplicative systematics, such that the corrected galaxy density field is estimated as
\begin{equation}
n_g^{\rm corr} = \frac{n_g^{\rm obs}}{1+E[\valpha]\mathbf{M}}.
\label{eq:mult1}
\end{equation}
For additive systematics one can apply the model of Eq.~\ref{eq:ols1} directly. 

There is however yet no formal way to distinguish between the additive and multiplicative character of systematics. As it will shown below they impact the observed galaxy field differently, particularly for moderate-to-high $\alpha$-amplitude values. This is the problem that we address in this work: we design and test a {\em hybrid } approach that attempts to handle and correct for both additive and multiplicative systematics.

\subsection{The hybrid approach}
\label{sec:hybrid}

Following the model given in Eq.~\ref{eq:smodel1}, here we outline a procedure that uses two different statistical techniques to solve for the $\vepsilon$ and $\vbeta$ vectors in that model. The first technique attempts to subtract the {\em} effective, additive total contribution of the systematics fluctuation vector $\delta \vMsys$ whose amplitude is given by $\vzeta$ ($\vzeta$ first appears in Eq.~\ref{eq:smodel2} but is defined right after Eq.~\ref{eq:dotMzeta}). Note that this $\vzeta$ amplitude contains contribution from both additive and multiplicative systematics, and is computed in step~1 and removed in step~3 below. Only once this is done, one can isolate the modulation of the variance of the resulting map induced by the purely multiplicative systematics (via the $\bar{n}_g\delta_g(\vbeta\cdot\delta \vMsys$) term in Eq.~\ref{eq:smodel2}). Isolating $\vbeta$ is the goal of the second statistical technique applied in steps 4 and 5 below. Once all possible purely multiplicative systematics have been corrected for, one attempts to correct for the additive ones, also in step 5 below. Our procedure relies on the existence of a set of mock galaxy catalogue whose angular statistical properties resembles those to be expected from the real, underlying galaxy population. We break this procedure in the following steps:
\begin{enumerate}
\item We first apply an OLS method on the observed galaxy map $n_g^{\rm obs} (\vnh)$. The OLS method provides an estimate for $\vzeta$, ${\rm E}[\vzeta]$, 
\begin{equation}
{\rm E}[\vzeta] = \frac{n_g^{\rm obs, t}(\vnh) \mathbf{C_M}^{-1} \delta\vM (\vnh)}{\delta\vM^t(\vnh)\mathbf{C_M}^{-1} \delta \vM (\vnh)},
\label{eq:OLS_est}
\end{equation}
with formal error given by
\begin{equation}
\sigma^2[\vzeta] = \frac{1}{\delta\vM^t(\vnh)\mathbf{C_M}^{-1} \delta \vM (\vnh)}.
\label{eq:OLS_err}
\end{equation}
In these equations, the matrix $\mathbf{C_M}$ is defined as the covariance matrix accounting for the angular correlation between different systematics template maps, $\mathbf{C_M}_{ij}:=\langle \delta M_i \delta M_j \rangle_{\vnh}$. The OLS assumes implicit Gaussian statistics for $n_g(\vnh)$ when solving for $\vzeta$  in $n_g^{\rm obs}(\vnh)=n_g(\vnh) + \vzeta\cdot\delta\vM (\vnh)$, and should yield an optimal estimate for $\vzeta$ in that case. In real galaxy surveys, it is preferred using a set of mock galaxy surveys when estimating the uncertainty in $\vzeta$, $\sigma [\vzeta]$, by computing the error distribution of the recovered $\vzeta$'s throughout all mocks. 
\item We denote {\it active} systematics those ones giving rise to some statistical evidence for leaving some imprint on $n_g^{\rm obs} (\vnh)$, via $|\zeta_i|>n_{\sigma} \sigma[\zeta_i]$, with $n_{\sigma} $ denoting some statistic threshold (e.g., $n_{\sigma} =3$)\footnote{Identifying which potential systematics are actually impacting the observed number of galaxies is a critical step, since neglecting any active template from the correction procedure (and/or including a non-active one) heavily impacts the quality of the corrected map. In \citet{weave+21} a ``elastic net" method is introduced to penalize the identification of too many templates as "active", while in our case we stick to frequentist statistical arguments on a set of mocks simulating the underlying galaxy density field, which allow us identifying which templates are {\em significantly} impacting the observed galaxy field.}.
\item We remove from the observed map our best estimate of the linear contamination induced by the active foregrounds,
\begin{equation}
n_g^{[1]} (\vnh) = n_g^{\rm obs} (\vnh) - \sum_i^{N_{\rm act}} {\rm E}[\zeta_i] \delta M_i (\vnh),
\label{eq:corrmap1}
\end{equation}
where the sum of $i$ runs over all active systematics templates. 
\item Since, according to Eq.~\ref{eq:smodel2}, the variance of $\bar{n}_g\delta_g$ is modulated linearly by $\vbeta\cdot \vMsys$ via the term $\bar{n}_g\delta_g(\vbeta\cdot \vMsys)$, and this remains so even after subtracting $E[\zeta_i]\delta M_i$ for the active systematics maps, the variance of the resulting map $n_g^{[1]} (\vnh) $ must still be affected by the multiplicative systematics. Hence we solve for the $\beta_i$'s by conducting a Monte Carlo Markov Chain (MCMC) minimization of the variance of the map
\begin{equation}
n_g^{[2]} (\hat{\beta_i},\vnh) = \frac{n_g^{[1]} (\vnh) }{\prod_i^{N_{\rm act}} (1+\hat{\beta}_i \delta M_i (\vnh)) }. 
\label{eq:corrmap2a}
\end{equation}
This minimization procedure yields a set of estimates for the $\hat{\beta}_i$'s that minimize the variance of $n_g^{[2]}$, hereafter denoted by ${\rm E}[\beta_i]$. We provide details of this minimization procedure in Appendix~\ref{sec:appB}, but we note here that only if $n_g^{[1]}(\vnh)$ is affected by multiplicative systematics the returned $E[\beta_i]$ values will be significantly different from zero.
\item Having solved for the $\beta_i$ values we can return, according to our model equation Eq.~\ref{eq:smodel1}, to the original observed map $n_g^{\rm obs}(\vnh)$ and correct for these multiplicative systematics, 
\begin{equation}
n_g^{[3]} (\vnh) =\frac{n_g^{\rm obs} (\vnh) }{\prod_i^{N_{\rm act}} (1+{\rm E}[\beta_i] \delta M_i (\vnh)) }.
\label{eq:corrmap2b}
\end{equation}
We apply the OLS again on the resulting map,
\begin{equation}
{\rm E}[\vepsilon_{\rm act}] = \frac{n_g^{[3],t} (\vnh) \mathbf{C_M}_{\rm act}^{-1} \delta\vM_{\rm act} (\vnh)}{\delta\vM_{\rm act}^t(\vnh)\mathbf{C_M}_{\rm act}^{-1} \delta \vM_{\rm act} (\vnh)},
\label{eq:corrmap3a}
\end{equation}
and remove these additive components in case they are significant, i,e., in case  $|{\rm E}[\epsilon_{{\rm act},i}]|>n_{\sigma} \sigma[\zeta_i]$, where the index $i$ runs through the active $N_{\rm act}$ systematics, and not all of the $N_{\rm act}$  actually fulfill this inequality. Denoting by $N_{\rm actA}$ the size of the subset of $N_{\rm act}$ templates verifying $|{\rm E}[\epsilon_{{\rm act},i}]|>n_{\sigma} \sigma[\zeta_i]$, we write the estimate for the corrected map of observed galaxy number density as 
\begin{equation}
n_g^{\rm corr} (\vnh)= n_g^{[3]} (\vnh)-\sum_i^{N_{\rm actA}} {\rm E}[\epsilon_{{\rm act},i}] \delta M_i (\vnh).
\label{eq:corrmap3b}
\end{equation}

From the latter equation, one can define a weight map given by 
\begin{equation}
{\cal W}(\vnh) = \frac{n_g^{\rm corr} (\vnh) }{n_g^{\rm obs}(\vnh)}.
\label{eq:weight1}
\end{equation}
This weight map can be interpreted as a re-scaling factor of the observed galaxy number density which corrects for the biases induced by imperfect observing conditions and any other possible astrophysical contaminants. 
\end{enumerate}

%_____________________
\begin{figure}
\hspace*{-0.5cm}
\includegraphics[scale=0.5]{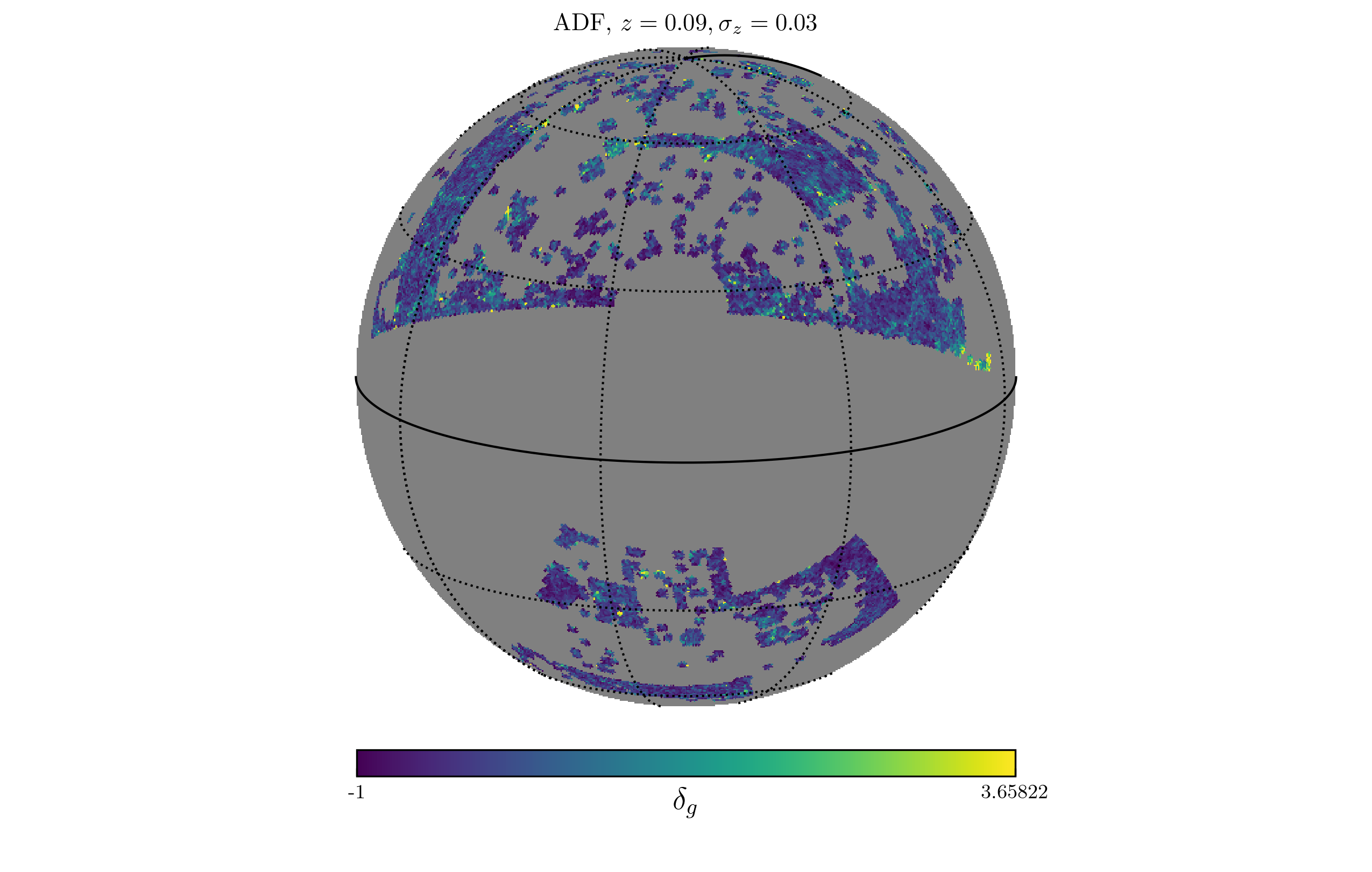}
\includegraphics[width=0.4\paperwidth]{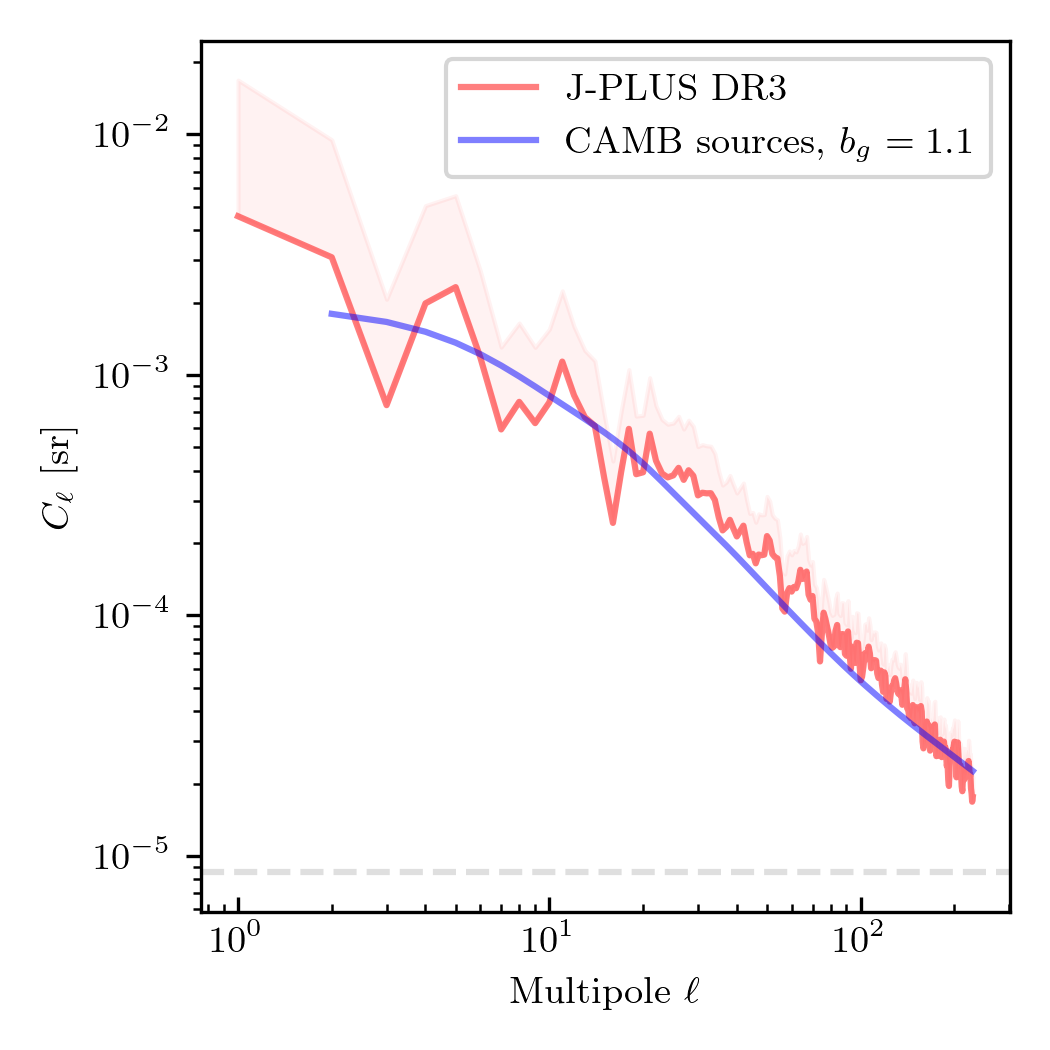}
\caption{ {\it (Top panel)}: Map of 2D angular density contrast ($\delta_g(\vnh)$) or angular density fluctuations (ADF) of J-PLUS DR3 galaxies  under a Gaussian redshift window centered upon $z=0.09$ and width $\sigma_z=0.03$. While some filamentary structure at high galactic latitudes can be hinted, at very low galactic latitudes, close to the galactic center, strong evidence for stellar contamination can also be seen. {\it (Bottom panel)}: Mask-corrected angular power spectrum of the map above (red line), after subtracting the shot noise term $C_{\ell}^{\rm SN}=1/\bar{n}_g$ induced by the finite number of galaxies (dashed, gray horizontal line). The term $\bar{n}_g$ denotes the average angular number density under this redshift shell. The shaded region above the red-line indicates the uncertainty region of its amplitude for every multipole $\ell$ under the approximation $\sigma^2 [C_{\ell}]= 2C_{\ell}/(2\ell+1)/f_{\rm sky}$, with $f_{\rm sky}\simeq 0.09$ the fraction of the sky covered by the galaxy survey. The theoretical prediction for the angular power spectrum of matter probes under the same Gaussian redshift kernel mentioned above and with linear bias $b_g=1.1$ is provided by the blue line, as computed by the Boltzmann code {\tt CAMB}. This computation accounts for the leading relativistic effects and a non-linear version of the 3D matter power spectrum. } 
\label{fig:ADFshellJPLUS}
\end{figure}
%-------------------------

%________________________________________________________________
\section{Results}
\label{sec:results}
\subsection{Results on J-PLUS-motivated mocks}
\label{sec:results_on_mocks}
We first test the performance of our hybrid algorithm on mocks motivated by the J-PLUS DR3 data. Our galaxy mocks will have angular clustering properties similar to those found in J-PLUS DR3 data, at least up to two-point statistics. They are generated as log-normal realisations of an angular power spectrum that has been previously extracted out of real J-PLUS data. 

In what follows we test the methodology on a single set of galaxy mocks that mimic the observed J-PLUS galaxy map after imposing a cut in the AUTO $r$-magnitude of $r<21$, {\tt odds}$>0.8$, and $p_{\rm star}<0.1$\footnote{The $p_{\rm star}$ parameter is obtained from the {\tt sglc\_prob\_star} entry in the data base, and provides the probability of a given object to be a star/compact source}. The {\tt odds} parameter constitutes a proxy for the quality of the redshift estimate provided by the photo-$z$s, and is defined as the fraction of the photo-$z$ PDF contained in the redshift range $[z_{\rm ML}-0.03\times (1+z_{\rm ML}),z_{\rm ML}+0.03\times (1+z_{\rm ML})]$, where $z_{\rm ML}$ is the most probable redshift given by the mode (or the maximum value) of the posterior photo-$z$ PDF \citep[see, e.g.,][]{AntonioHC_phz1}. We build a 2D galaxy density map after weighting each galaxy falling on a sky pixel by a Gaussian weight $W\propto \exp{ [-(z_{\rm ML}-z_0)^2/(2\sigma_z)^2] }$, with $z_{\rm ML}$ the corresponding most probable redshift of the galaxy, and where the central redshift has been taken to be $z_0=0.09$ and $\sigma_z=0.03$\footnote{These are just typical values for $z_0$, $\sigma_z$ for J-PLUS DR3 tomographic analyses.}. We stress this is an arbitrary choice that gives rise to a galaxy density map whose angular power spectrum is used to generate our galaxy mocks. The goal here is testing the systematic-correction methodology on a given shell (that in practice is applied on all redshift shells under analysis), and we defer the detailed and systematic clustering analysis of J-PLUS galaxy samples for future work.

%_____________________
\begin{figure}[hbt!]
\includegraphics[width=0.4\paperwidth]{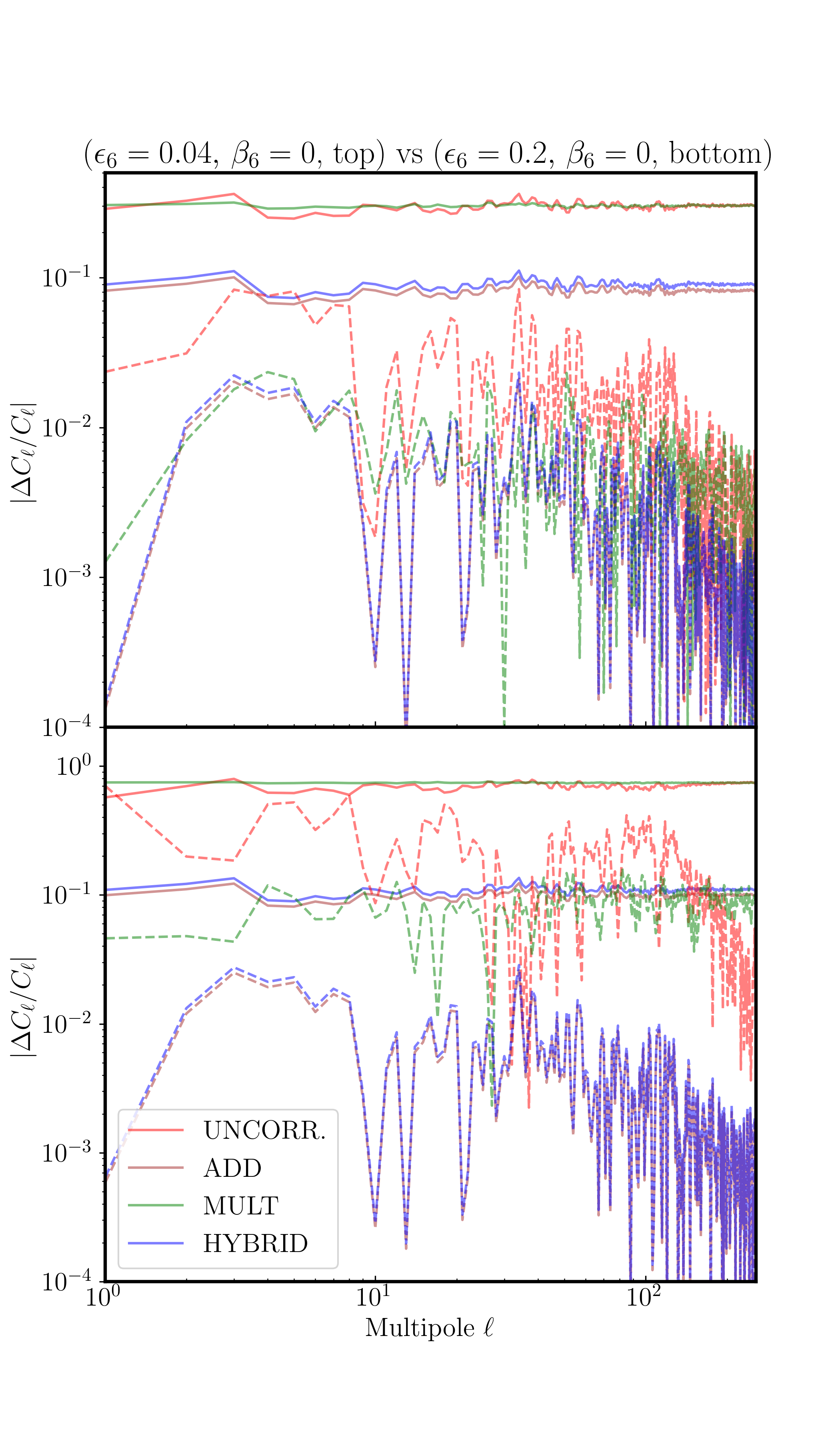}
\caption{Comparison of the relative error in the recovery of the angular power spectrum under a single, additive contaminant. The top and bottom panels display two cases of low ($\epsilon=0.04$, $\beta=0$) and high ($\epsilon=0.2$, $\beta=0$) contamination, respectively. Solid lines assume no extra information about the average galaxy density $\bar{n}_g$, which is estimated at systematic removal. Dashed lines, on the contrary, assume that $\bar{n}_g$ is provided with arbitrary precision. For clarity reasons, the results of the hybrid approach have been boosted by a factor $1.1$, otherwise they cannot be discerned from the outcome of the additive approach. }
\label{fig:means1}
\end{figure}
%-------------------------

For illustration purposes, in Fig.~\ref{fig:ADFshellJPLUS} we display a map of galaxy density for the quoted Gaussian redshift shell centred upon $z_0=0.09$ and $\sigma_z=0.03$ (top panel), together with its angular power spectrum (bottom panel). In this panel, we compare the raw angular power spectra from our data with a theoretical prediction provided by the Boltzmann code {\tt CAMB} \citep{camb_ref} under a standard flat $\Lambda$CDM cosmology\footnote{The adopted values for the relevant cosmological parameters are $\Omega_b h^2=0.022$, $\Omega_c h^2=0.122$, $h=67.5$, $A_s=2\times 10^{-9}$, $n_S=0.965$ for the baryonic physical density, cold dark matter physical density, reduced Hubble constant, amplitude of the primordial scalar power spectrum, and spectral index of the scalar power spectrum, respectively. All these values are compatible with the latest data release of the {\it Planck} experiment \citep{planck2018}. }, after taking a linear bias $b_g=1.1$ and the default non-linear approximation for the underlying 3D power spectrum. The angular power spectrum can be written as a  mere projection of the 3D galaxy power spectrum into multipole $\ell$ space,
\begin{equation}
C_{\ell} =\frac{2}{\pi} \int dk \, k^2 \,P_{g}(k) |\Delta_{\ell}^g(k)|^2,
\label{eq:cls2pk}
\end{equation}
where the $\Delta^g_{\ell}(k)$ are transfer functions that contain information about the bias of the galaxy population and the amplitude of their peculiar, radial velocities \citep[see, e.g.,][]{camb_ref}. At the same time, the amplitude and shape of the galaxy 3D power spectrum ($P_{g}(k)$) are sensitive to the physics of the inflationary epoch of the universe and the epoch of matter-radiation equality. Thus, the $C_{\ell}$'s of a density map of any given galaxy population carry imprints of extremely interesting cosmological and astrophysical physics, and this motivates the need for their accurate measurement and interpretation.
 
\subsubsection{The impact of the template monopole}

%_____________________
\begin{figure*}
\includegraphics[width=0.9\paperwidth]{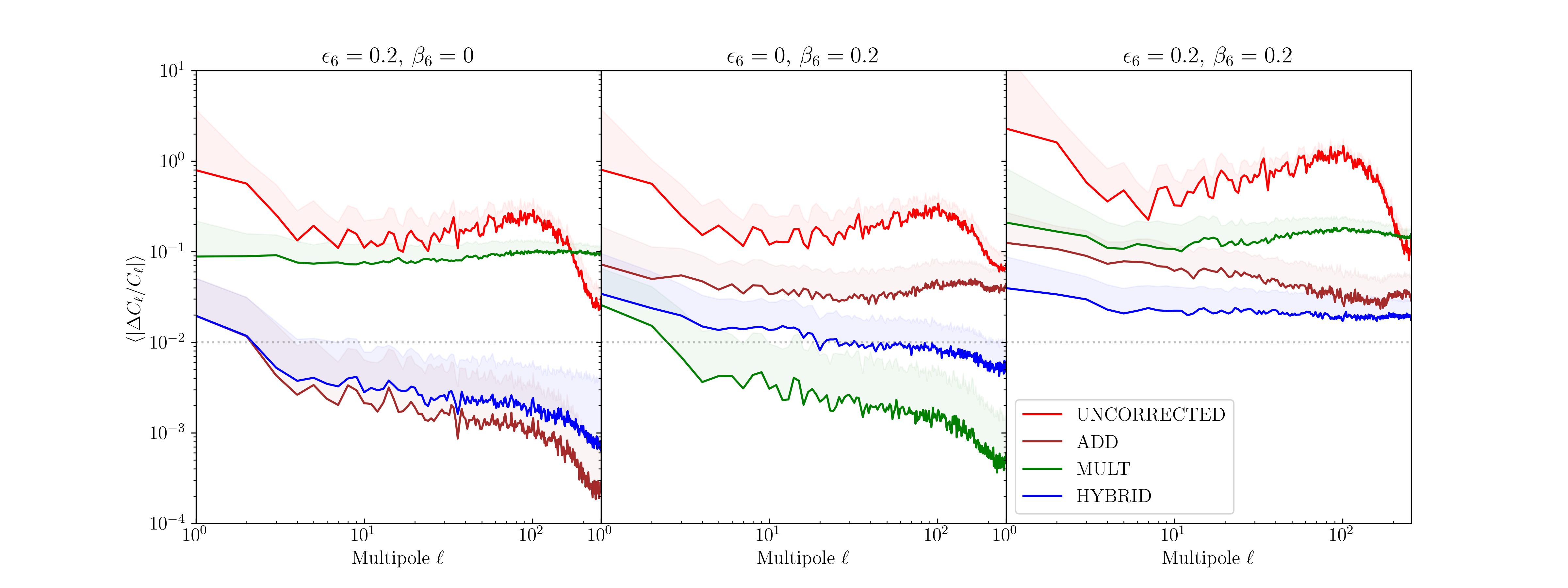}
\caption{Recovery of angular power spectrum after biasing 200 log-normal galaxy mocks with the {\tt fwhmg} systematics template. This bias may be either additive ($\epsilon_6=0.2,\,\beta_8=0$, left panel), multiplicative ($\epsilon_6=0,\,\beta_8=0.2$, middle panel), or both additive and multiplicative ($\epsilon_6=0.2,\,\beta_6=0.2$, right panel). }
\label{fig:onetemp6}
\end{figure*}
%-------------------------

%_____________________
\begin{figure*}
\includegraphics[width=0.9\paperwidth]{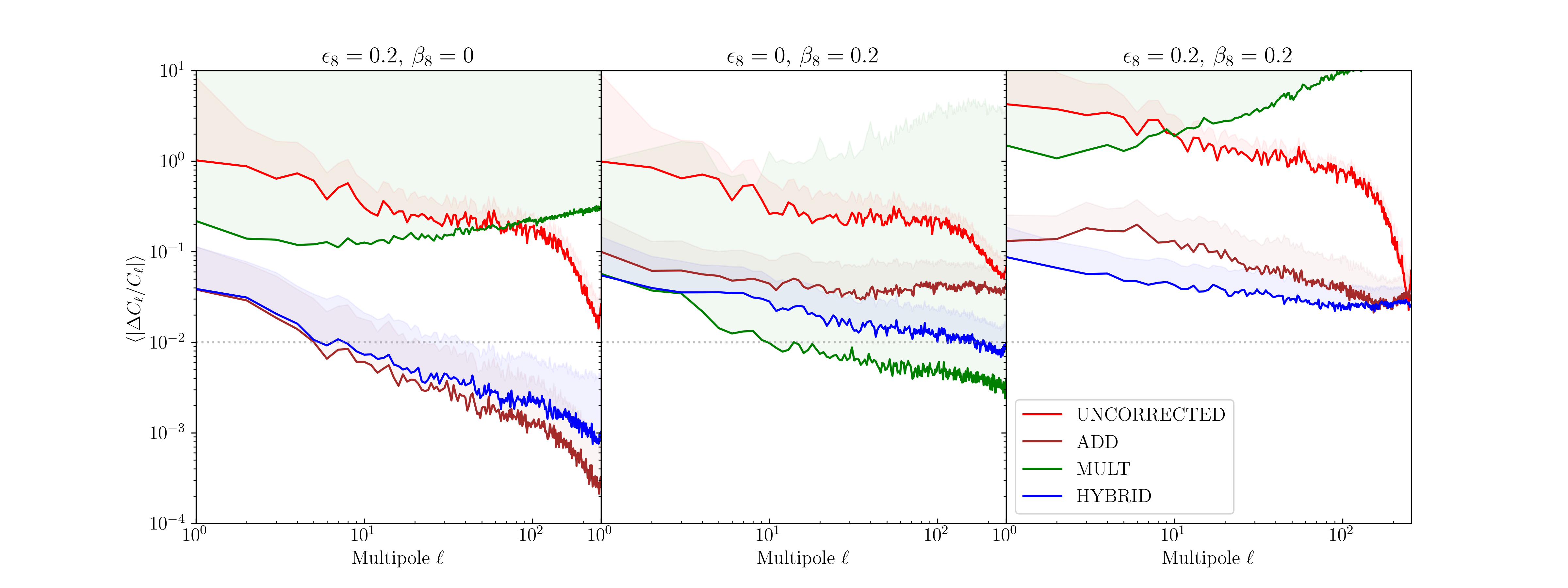}
\caption{ Same as in Fig.~\ref{fig:onetemp6}, but for the {\tt deep2fwhm} template. }
\label{fig:onetemp8}
\end{figure*}
%-------------------------

We have seen in Sect.~\ref{sec:model} that the monopole/angular average of systematics templates only impacts the observed galaxy density if systematics are additive, and that multiplicative systematics do not bias the galaxy map monopole at linear order in $\epsilon,\,\beta$. So we next address a simple example where we consider a single, additive systematics template, which in practice corresponds to {\tt fwhmg} (referring to the effective fwhm after assuming a Gaussian shape of the point sources falling in a given pixel). In this case we have $n_g^{\rm obs}(\vnh) = \bar{n}_g(1+\delta_g[\vnh]) + \alpha M(\vnh)$. The angular average (or monopole) of this template map is positive ($\bar{M}>0$), so if added to the underlying galaxy density field, it will bias (increase) its monopole or angular average: $\langle n_g^{\rm obs}(\vnh)\rangle_{\vnh}=\bar{n}_g + \alpha \bar{M}$. Since the angular density contrast is built after normalising by this angular average,  
\begin{equation}
\delta_g^{\rm obs} (\vnh) = \frac{n_g^{\rm obs}(\vnh) }{\langle n_g^{\rm obs}
(\vnh)\rangle_{\vnh}} -1,
\label{eq:delta_g_obs}
\end{equation}
a wrong normalization by the average galaxy density results in a wrong amplitude for the density contrast, and most of the cosmological statistics estimated thereof.

One could argue that if this systematics contribution is correctly identified as an additive one, then this bias induced by the contaminated average galaxy density can be corrected by, e.g., the OLS method outlined above. However, all current methodology on systematics removal/correction relies on the {\em spatial} comparison of systematics templates with the observed galaxy density field. That is, all methods are {\em only} sensitive to spatial/angular variations of both the templates and the galaxy field, but they are not directly sensitive to a non-zero contribution to the observed average galaxy density. If a systematics template was constant on the sky, without angular variations, it would not be possible to estimate its contribution to the average/monopole of the observed galaxy density map. In the case of additive systematics, the contribution to the observed mean galaxy density can only be corrected if the model $n_g^{\rm obs}(\vnh) = \bar{n}_g(1+\delta_g[\vnh]) + \alpha M(\vnh)$ applies {\rm strictly}, that is, if both the angular average (or monopole, $\bar{M}$) and the spatially varying part ($\delta M(\vnh)$) of the systematics template $M (\vnh)$ contribute to the observed galaxy number density under the same constant amplitude $\alpha$. We sense this may be a strong assumption, which we would like to relax given the non-trivial way systematics may be impact the observed number density of galaxies.

We prefer working instead in a second scenario where the average galaxy density $\bar{n}_g$ is estimated from purity measurements in the galaxy samples. We thus assume that the survey can provide an unbiased estimate of $\bar{n}_g$ that accounts for the presence of all possible contaminants (like stars or high-redshift quasars in the case of, e.g., a low $z$ galaxy survey as J-PLUS). Under this assumption we can ignore the constant terms in Eq.~\ref{eq:smodel3}, which is equivalent to dropping all template monopoles ($\bar{M}_i=0$) in our templates. We then handle the angular variations of the observed galaxy density field, $\delta n_g^{\rm obs}(\vnh) =  n_g^{\rm obs}(\vnh) - \langle n_g^{\rm obs}(\vnh) \rangle_{\vnh}$, and after correcting for systematics we normalize the corrected $\delta n_g^{\rm corr} (\vnh)$ field by the purity estimate of $\bar{n}_g$. The uncertainty in this estimate will be propagated to the uncertainty of the estimated galaxy density contrast field, $\delta^{\rm corr}_g(\vnh)$.

A simple but illustrative example for these two scenarios is provided in both panels of Fig.~\ref{fig:means1}, where we study the impact of adding the {\tt fwhmg} spatial template to a mock log-normal realization of a galaxy field resembling the reference J-PLUS galaxy map centred upon $z_0=0.09$. Since we know the map before and after contamination by the template, we can compare the initial angular power spectrum with the contaminated one, and with the ones derived from the ``corrected" maps. The top panel corresponds to a case of mild to low additive contamination ($\epsilon=0.04$, $\beta=0$), whereas the bottom one refers to ($\epsilon=0.2$, $\beta=0$). The first scenario where there is no prior knowledge on $\bar{n}_g$ is displayed by solid lines in the plot, whereas the second one, where it is assumed that the error on $\bar{n}_g$ is negligibly small, is portrayed by the dashed lines. In the former the red lines, corresponding to the uncorrected angular power spectrum, shows the error/amplitude scale owed to the contribution of the additive systematics to the global monopole in the galaxy map ($\bar{M}$). While the multiplicative approach (green solid lines) cannot account for this bias by construction (and thus yields roughly the same relative error as the uncorrected case), the additive and hybrid methods manage to partially correct for the amplitude error, decreasing it to 30\%--10\% of its original amplitude.  In the second scenario relative errors on the angular power spectrum show a different behaviour: even the uncorrected case (red dashed lines) shows structure versus multipole $\ell$, and in this case the multiplicative approaches manages to correct for part of the error, although they are typically outperformed by the additive and hybrid approaches (which give virtually indistinguishable results, reason for which the curves for the hybrid approach have been boosted by 10\%). For low contamination (top panel) the difference between the multiplicative approach on the one hand, and the additive and hybrid ones on the other is very subtle, particularly at low multipoles. When the amount of contamination becomes more relevant (bottom panel), the additive and hybrid methods outperform the multiplicative one more clearly. 

As mentioned above, in what follows in this work $\bar{n}_g$ estimates will be assumed to be provided externally to the process of systematics removal/amelioration. We thus factorize out the uncertainty associated to $\bar{n}_g$ that projects into an amplitude uncertainty of the density contrast power spectrum, and neglect the impact of the template monopoles ($\bar{M}_i$s): the comparison among the three different approaches (additive, multiplicative, and hybrid) will be conducted in the scenario depicted by dashed lines in Fig.~\ref{fig:means1}.

\subsubsection{Results for a single systematics template}
\label{sec:results_on_single_mock}

In this subsection we study the performance of the three approaches when contaminating a set of 200 mocks inspired in our J-PLUS DR3 reference galaxy map with a single template. Since only one template will impact the observed galaxy density field, we opt for a relatively high value of $\epsilon,\,\beta=0.2$: in the linear or quasi-linear scale regime we are working at the density contrast rms should lie at the $\lesssim 0.3-0.1$ level. In Fig.~\ref{fig:onetemp6} we show the results after including the systematics template tracking {\tt fwhmg}, which is sensitive to the effective seeing in each tile. In Fig.~\ref{fig:onetemp8} we repeat the same exercise but for the template {\tt depth2fwhm}, which constitutes a proxy for the effective photometric depth in each pixel of the footprint, and shows different angular clustering properties to those of the {\tt fwhmg} template. In both figures, left panels consider the purely additive case ($\epsilon=0.2,\,\beta=0$), the middle panels the purely multiplicative case ($\epsilon=0,\,\beta=0.2$), and the right panels the hybrid case with both additive and multiplicative contamination ($\epsilon=0.2,\,\beta=0.2$). The color coding refers to the different cleaning procedures being applied, and in all cases solid lines are showing the median of the {\em absolute value} of the relative error of the recovered angular power spectrum multipoles, $|\Delta C_{\ell}/C_{\ell}|$. The shaded regions above the solid lines are limited, from above, by the sum of the median of $|\Delta C_{\ell}/C_{\ell}|$ plus its rms throughout the mocks, for every multipole $\ell$. These regions thus provide a hint on the (half) width of the distribution of $|\Delta C_{\ell}/C_{\ell}|$ values around their median given by the solid lines, and provide a measure of the uncertainty associated to each cleaning procedure.

Comparing results displayed in Fig.~\ref{fig:onetemp6} and Fig.~\ref{fig:onetemp8} shows qualitative agreement between the two systematics templates. For both templates, the additive and multiplicative approaches yield best results in terms of the median of $|\Delta C_{\ell}/C_{\ell}|$ for pure additive and multiplicative contamination, respectively. However, the hybrid method is second best, following from not far behind, and when the contamination is hybrid (both additive and multiplicative systematics contribution), the hybrid method outperforms the other two. Interestingly, for the {\tt deep2fwhm} template in the purely multiplicative case (middle panel of Fig.~\ref{fig:onetemp8}, $\epsilon=0,\,\beta=0.2$), one could argue that actually the multiplicative approach is not the optimal one, despite the fact of yielding the lowest curve for the median $|\Delta C_{\ell}/C_{\ell}|$ versus $\ell$. The very extended character of the green shaded region is pointing to very large rms values in the distribution of the corrected angular power spectrum multipoles for this approach. This can also be seen in the additive and hybrid cases. This is due to the presence of large amplitude regions in this template, which can make the multiplicative approach diverge when certain values of $\beta$ take the correction map $1/(1+\beta \delta M(\vnh))$ to arbitrarily large values. This situation is, by design, avoided by the minimum variance requirement in the hybrid approach. 

\subsubsection{Results for multiple systematics templates }
\label{sec:results_on_multiple_mocks}

We next compare the performance of the additive, multiplicative, and hybrid methods when a set of several, partially correlated systematics templates are biasing the observed number of galaxies. Our total set consists of 14 different templates, and some of them show similar structures on the sky, such as the template {\tt stars} and {\tt ebv} or {\tt ebvPlanck}, which reveal more structure in and at the vicinity of the Galactic plane. In this occasion, we choose to bias the 200 log-normal mocks with the templates of indexes ranging from 5 to 11, namely {\tt ncombined}, {\tt fwhmg}, {\tt m50s}, {\tt depth2fwhm}, {\tt depth3as}, {\tt ebv}, and {\tt ebvPlanck}. These 7 templates can be grouped in four separate groups of templates, such that templates within the same group show clear similarities, but members of different groups show no obvious correlation. 

Since our methodology is blind to which systematics are actually modifying the observed number of galaxies, this set of analyses will quantify realistically the impact of misidentifying active systematics in our pipelines. Failing to label as active a systematics template that is actually impacting the observed galaxy map, or attempting to correct one or more templates that in reality are not biasing observations will have an impact in the error budget of the recovered angular power spectra. Typically when one template is identified as active, then all other systematics templates belonging to the same group of correlated templates will also be tagged as active.

\begin{figure*}
\includegraphics[width=0.9\paperwidth]{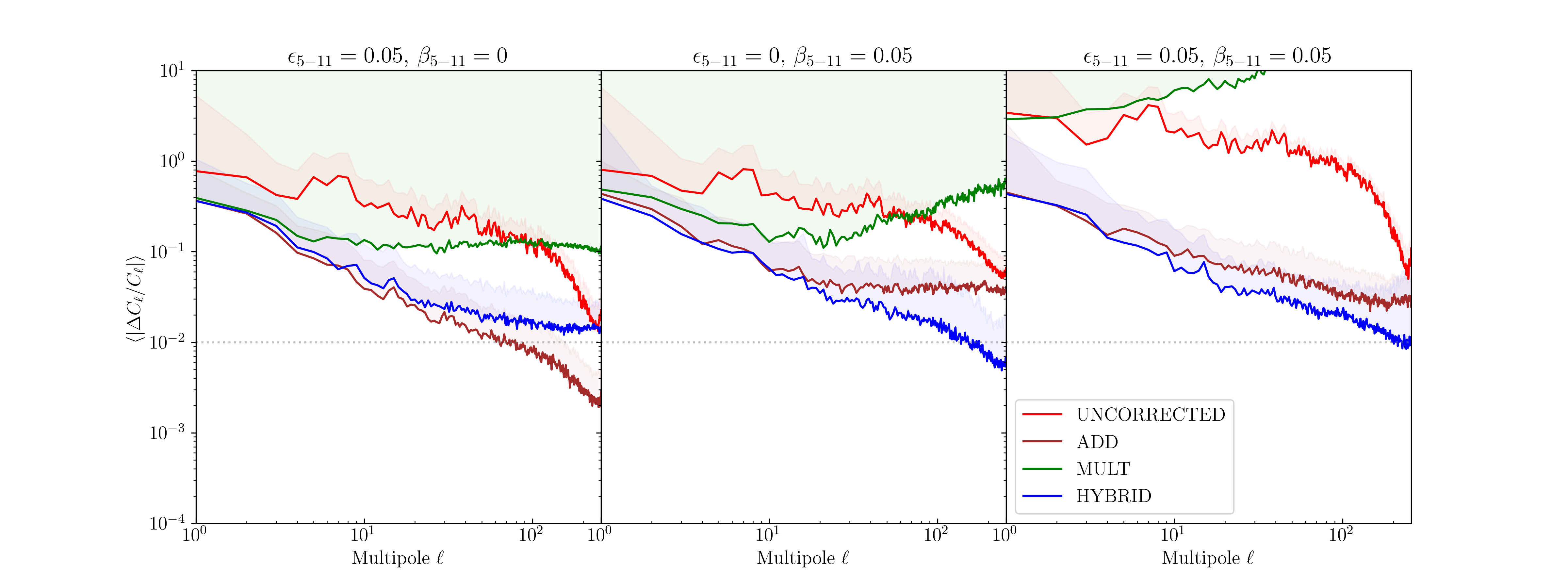}
\caption{ Recovery of angular power spectrum after biasing 200 log-normal galaxy mocks with templates with indexes ranging from 5 to 11. We consider a relatively large amount of contamination per systematics template: $\epsilon_{5-11}=0.05,\,\beta_{5-11}=0$ for the purely additive scenario in the left panel, $\epsilon_{5-11}=0,\,\beta_{5-11}=0.05$ for the multiplicative one in the middle panel, and $\epsilon_{5-11}=0.05,\,\beta_{5-11}=0.05$ in the hybrid scenario shown in the right panel. Shaded areas display the upper half of the rms uncertainty region around the median values given by the solid lines.  
}
\label{fig:sevtemps_0p05}
\end{figure*}

In our first exercise we examine the outcome of the three approaches after including moderate to high contamination on the mocks. We do so by considering an amplitude of $0.05$ in the $\epsilon,\,\beta$ parameters for templates 5 to 11 (see Fig.~\ref{fig:sevtemps_0p05}). As in previous exercises, the total bias in the angular power spectrum multipole after contamination typically decreases from $\sim 100\%$ at low multipoles down to $\sim 10\%$ on small scales (large multipoles). We find in Fig.~\ref{fig:sevtemps_0p05} that, for this configuration, the multiplicative approach is unstable in all three scenarios (additive/left panel, multiplicative/middle panel, and hybrid/right panel) considered. The hybrid approach performs better than the additive one in the multiplicative and hybrid exercises, and the reverse situation applies for the additive exercise (left panel). At low multipoles the additive and hybrid approaches yield very similar errors, and differences between these two methods become more relevant on the small scales. The improvement with respect to the uncorrected case (given by the red line) is actually higher for higher contamination levels: in the right panel we can see that at multipole $\ell\simeq 100$ the hybrid method residual level is about/below 2\%, while the bias in the uncorrected case at that same multipole amounts to $\sim 100\%$, thus yielding an improvement of a factor of $\sim 50$. This factor is noticeably smaller when the total contamination level is also lower, as shown in the left and middle panels of the same Fig.~\ref{fig:sevtemps_0p05}: for the hybrid approach it lowers to $\sim 10,\,20$ for the pure multiplicative ($\epsilon_{5-11}=0$) and additive ($\beta_{5-11}=0$) cases, respectively.

\begin{figure*}
\includegraphics[width=0.9\paperwidth]{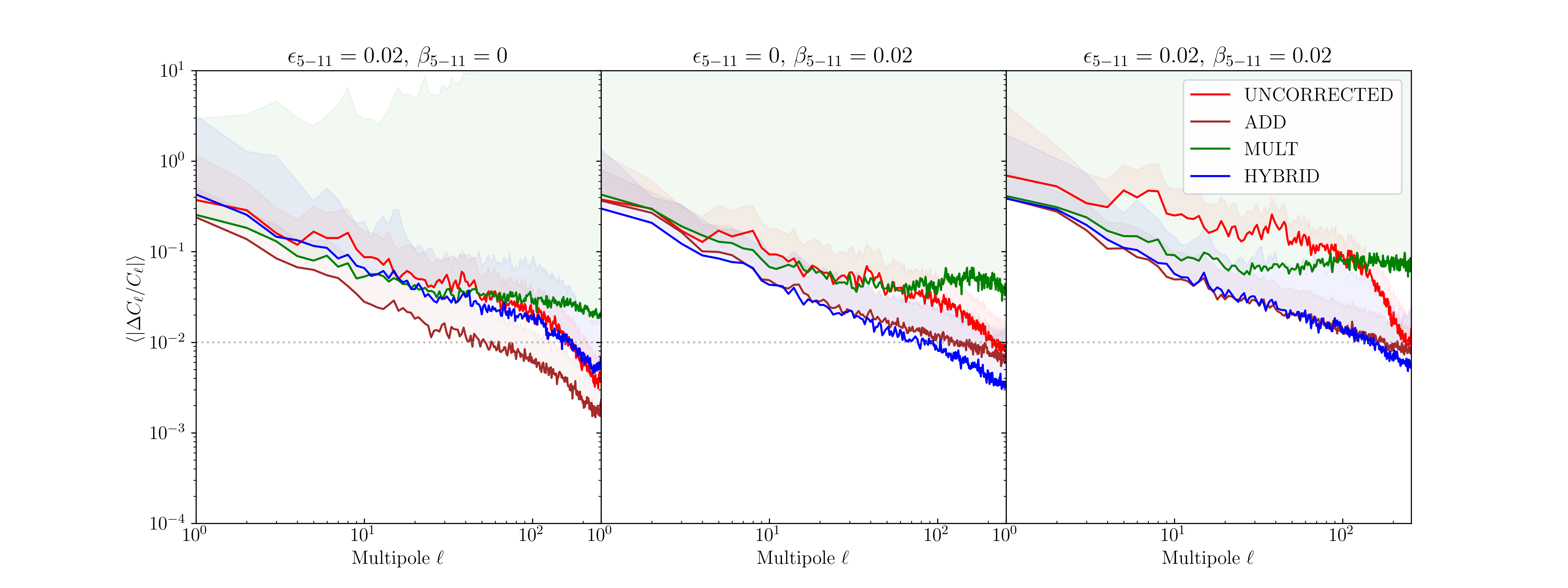}
\caption{ Same as in Fig.~\ref{fig:sevtemps_0p05}, but for lower contamination levels ($\epsilon_i,\beta_i=0.02$).}
\label{fig:sevtemps_0p02}
\end{figure*}

These results seem to suggest that there exists a floor in error/residuals level which the methods cannot easily reach. This partially motivates the exercise shown in Fig.~\ref{fig:sevtemps_0p02}, where systematics templates from 5 to 11 are given an amplitude of $0.02$ when acting on the galaxy mocks. This naturally results in a corresponding lower bias level on the angular power spectrum multipoles, and yet a lower improvement factor when comparing to the uncorrected cases given by the red lines. In the purely multiplicative and hybrid scenarios of the middle and right panels the additive and hybrid approaches (brown and blue lines, respectively) lie very close (although their shaded regions may differ). In the middle panel the residual bias of the corrected power spectra for both additive and hybrid approaches is typically a factor of only $\sim 3$ below the uncorrected case, while this factor becomes close to $\sim 10$ in the hybrid scenario of the right panel (for a wide range of multipoles). But in both panels the level of the residuals is very similar, pointing to a limiting residual/bias level, presumably determined by the finite statistics associated to the limited extension of the survey's footprint (which is covering, at best, $4\pi$ sr, but only $\sim 3\,000$~sq.deg. in our simulated mocks). We shall elaborate further this point in Sect.~\ref{sec:discussions}.

The additive scenario of the left panel shows a significant difference between the additive and hybrid approaches. The results in this panel, when put in context with the other two panels of this figure, suggest that, for this template set and low contamination levels (i.e., for low statistical significance of the OLS outputs), the use of the additive algorithm should be considered along with the hybrid one. We shall address this comparison below in Sect.~\ref{sec:discussions}.

%_____________________
%-------------------------

%_____________________
%-------------------------

\subsection{Results on a redshift shell of J-PLUS DR3}
\label{sec:results_on_JPLUSDR3}

%_____________________
\begin{figure*}
\includegraphics[width=0.8\paperwidth]{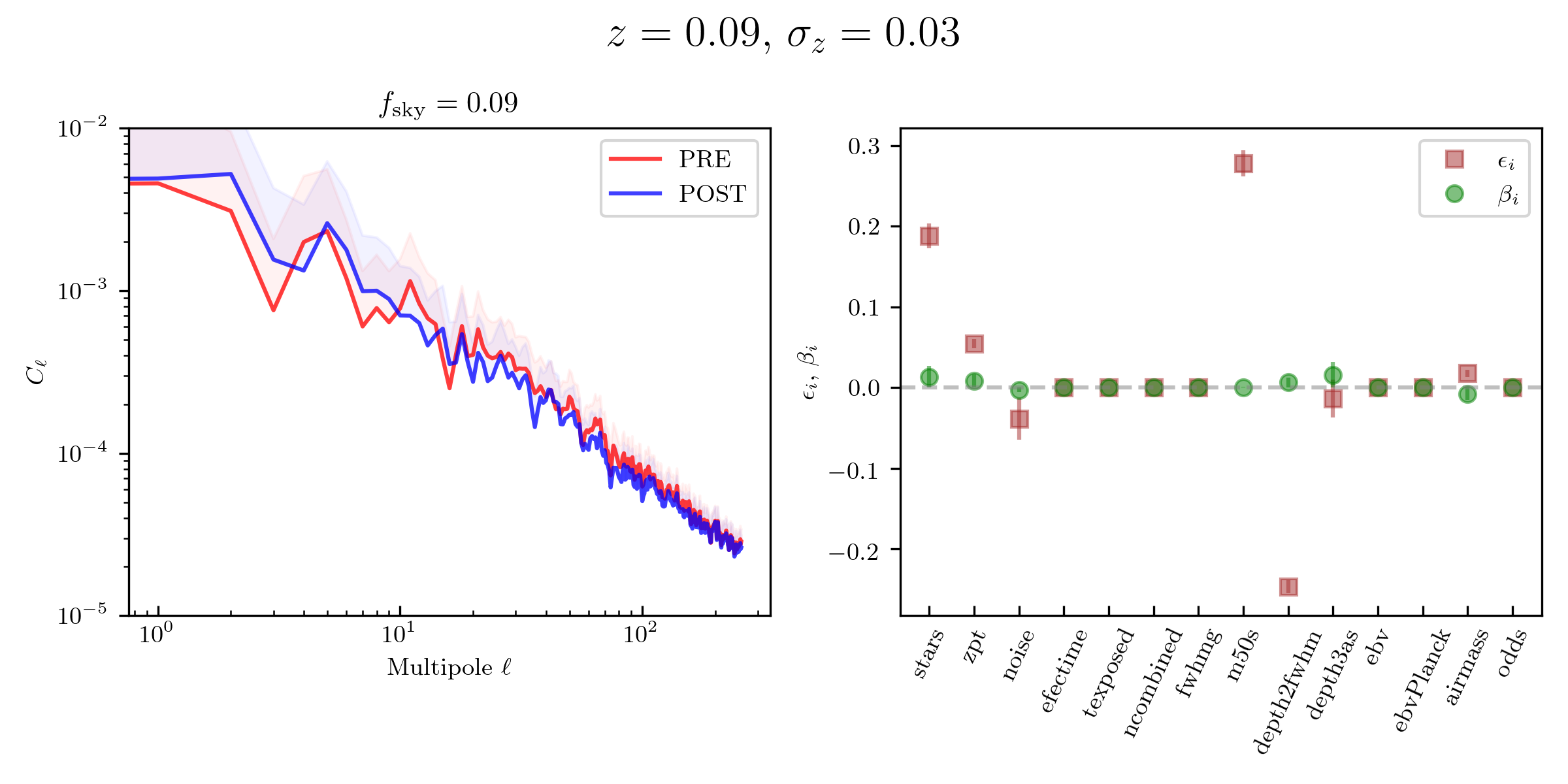}
\caption{ Application of our systematics-removal methodology on a J-PLUS DR3 galaxy sample weigthed under a Gaussian redshift shell centered upon $z=0.09$ and of width $\sigma_z=0.03$. {\it (Left panel:)} Angular power spectrum of original (red) and {\it corrected} (blue) galaxy density map. Shaded areas display the upper half of the rms region, just as in Fig.~\ref{fig:ADFshellJPLUS}.  {\it (Right panel:)} Contribution from the different systematics templates considered in this work, both in their additive ($\epsilon_i$) and multlicative ($\beta_i$) version.}
\label{fig:eps_beta_DRshell}
\end{figure*}
%-------------------------

In this subsection we briefly describe the application of our hybrid methodology on the real J-PLUS DR3 galaxy density map we have used so far as reference. In the right panel of Fig.~\ref{fig:eps_beta_DRshell} we show the recovered value of $\epsilon_i$ and $\beta_i$ for all systematics templates under consideration. We can see that there exists higher evidence for the {\tt stars} and {\tt zpt} systematics templates in the observed galaxy density map,  Note as well that the impact of {\tt m50s} and {\tt depth2fwhm} is apparently very significant: however, the fact that they are of opposite sign, and that another systematics template that is also probing the photometric depth  {\tt depth3as}) yields no statistically significant output, points to joint negligible impact of these templates. This is a typical expression of the degeneracy of different but correlated angular templates: while some of them may seem to individually (and significantly) impact the observed galaxy density map, one should observe their joint/combined contribution. It is also noticeable that, for this galaxy density map, the variance of the original density map seems not to be heavily impacted by the systematics since most of the $\beta_i$ measurements are very close to zero. 
We have checked that most of the contribution in the change of the recovered angular power spectrum (left panel) is actually due to the impact of the first two templates ({\tt stars} and {\tt zpt}), and that the recovered map under the assumption of additive systematics is very close to the one rendered by the hybrid approach. \\

In a dedicated study \citep{chm-jplus-tomoDR3} we thoroughly study the impact of all the systematics templates under varying width Gaussian redshift shells from $z=0.03$ up to $z=0.39$, particularly on their corresponding angular power spectra.

\subsection{Angular redshift fluctuations in the presence of systematics}

We next briefly address the behavior of angular redshift fluctuations \citep[hereafter ARF,][]{arf_letter1,adal_chm} in the presence of systematics biasing the observed number density of galaxies. ARF express the angular anisotropies of the redshifts of matter probes after projecting them under a finite redshift shell $W(z)$. As shown in \citet{adal_chm}, there exist at least two different definitions of the ARF that yield the same expression at linear level of cosmological perturbation theory:

\begin{eqnarray}
\label{eq:dz_I}
 (\delta z)^I (\vnh) & = & \frac{\sum_{j\in \vnh} z_j\,W(z_j)}{ \sum_{j\in \vnh} W(z_j) } - \bar{z} \,\,; \\
\label{eq:dz_II}
  (\delta z)^{II} (\vnh) & = & \frac{\sum_{j\in \vnh} (z_j-\bar{z})W(z_j)}{\langle \sum_{j\in \vnh} W(z_j) \rangle_{\vnh}},
\end{eqnarray}
with $z_j$ the observed redshift of the $j$-th galaxy, $\bar{z} := \ \sum_j z_j \, W(z_j) / \sum_j W(z_j)$ the average redshift under the redshift shell $W(z)$, and $\langle ... \rangle_{\vnh}$ denoting the angular average throughout the entire survey footprint. 

As just mentioned above, when writing both observables in terms of cosmological perturbations of the matter density and peculiar radial velocity fields, one arrives to the same, identical expression at leading (linear) order, \citep{arf_letter1}. In that work only the (dominant) density and radial velocity gradient terms are considered. We next study how systematics affect those two different approaches to measure the ARF. 

If one neglects systematics impacting the observed redshifts and restricts to those biasing the observed number of galaxies, then it is clear from the above expressions that purely multiplicative systematics will not bias the estimator in Eq.~\ref{eq:dz_II} {\em if} they do not change under $W(z)$, since they will contribute to exactly the same in both the numerator and the denominator of Eq.~\ref{eq:dz_II}. That is, if $\bar{n}_z(z) := dN/dz(z)$ refers to the average redshift galaxy density, then we find that
\[
(\delta z)^{II}(\vnh) = \frac{\int dz W(z)\,z\, (1+\beta(\vnh))\bar{n}_z(1+\delta_g (z,\vnh))}{\int dz W(z)(1+\beta(\vnh)\bar{n}_z(1+\delta_g (z,\vnh))} 
\]
\[
\phantom{xxxx}
 = \frac{(1+\beta(\vnh))\int dz W(z)\,z\, \bar{n}_z(1+\delta_g (z,\vnh))}{(1+\beta(\vnh)) \int dz W(z)\bar{n}_z(1+\delta_g (z,\vnh)}
\] 
\begin{equation}
\phantom{xxxx}
  = \frac{\int dz W(z)\,z\, \bar{n}_z(1+\delta_g (z,\vnh))}{\int dz  W(z)\bar{n}_z(1+\delta_g (z,\vnh))}.
\label{eq:dzII_Msyst}
\end{equation}
I.e., we end up with the same expression as in the no-systematics case.
However, for additive systematics that remain constant under $W(z)$ we do not find that same (fortunate) cancellation:
\[
(\delta z)^{II}(\vnh) = \frac{\int dz W(z)\,z\,\bar{n}_z(1+\delta_g (z,\vnh)+\epsilon(\vnh))}{\int dz W(z)\bar{n}_z(1+\delta_g (z,\vnh)+\epsilon(\vnh))} 
\]
\begin{equation}
\phantom{xxxxxxxxxxx}
 = \frac{\epsilon(\vnh) \bar{N}\bar{z} + \int dz W(z)\,z\, \bar{n}_g(1+\delta_g (z,\vnh))}{\epsilon(\vnh) \bar{N} + \int dz W(z)\bar{n}_z(1+\delta_g (z,\vnh))},
\label{eq:dzII_Asyst}
\end{equation}
with $\bar{N}:=\int dz\,\bar{n}_z\,W(z)$. More generally, both additive and multiplicative systematics will {\it a priori} bias the estimate in Eq.~\ref{eq:dz_I}, so one can state that Eq.~\ref{eq:dz_II} introduces a more robust ARF estimator than Eq.~\ref{eq:dz_I} with respect to multiplicative systematics that remain constant under the redshift shell. We have confirmed these prediction using our log-normal mocks and systematics templates.

However, one must have present that Eq.~\ref{eq:dz_II} contains the actual galaxy number along $\vnh$ in the denominator, and this may introduce instabilities and biases (caused by shot noise associated to the tracer number) for surveys with low galaxy number densities under $W(z)$. It is also important to remember that the fully relativistic ARF computation given in \citet{adal_chm} applies only to Eq.~\ref{eq:dz_I} and not to Eq.~\ref{eq:dz_II}. Thus, the use of either estimator will depend upon the particular case under study.

%________________________________________________________________
\section{Discussion}
\label{sec:discussions}

%_____________________
\begin{figure*}
\includegraphics[width=0.9\paperwidth]{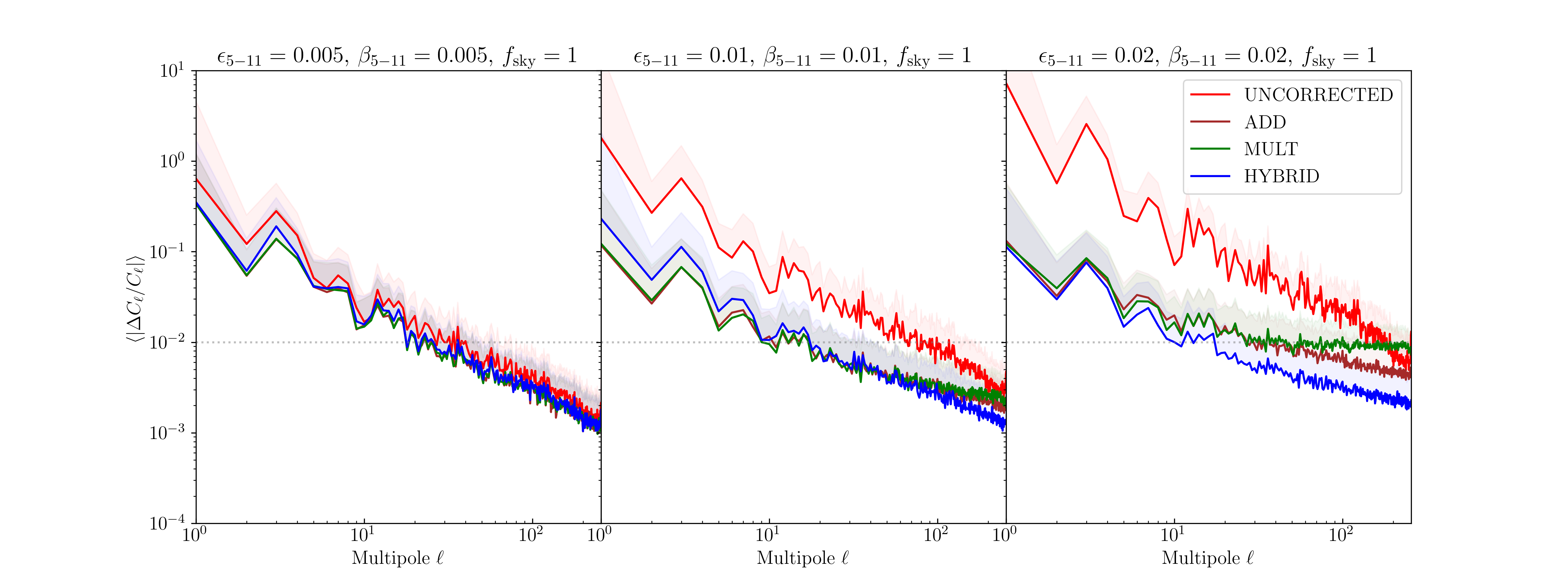}
\caption{Extension of our template set to the full sky, and comparison of the three methods under different levels of impact under the hybrid (both additive and multiplicative) scenarios. Shaded areas display the upper half of the rms uncertainty region around the median values given by the solid lines.  }
\label{fig:allskytsts}
\end{figure*}
%-------------------------

Our work has shown that additive and multiplicative systematics impact the observed field of galaxy density in different ways, and that, precisely for this reason, it is possible to distinguish between the two families of potential systematics. As first shown by \citet{HenriqueXabier}, the minimization of the variance has proven to be a powerful tool to identify multiplicative systematics, and a {\em hybrid} algorithm may be devised with the standard OLS approach to provide a method that simultaneously searches for both additive and multiplicative systematics (and even a manifestation of both characters for the same source of systematics). 

Confusing an additive with a multiplicative systematic (and vice versa) reflects on larger errors of the recovered angular power spectrum under moderate- to high-levels of contamination. In these circumstances, our hybrid method of systematics correction proves to be significantly more powerful in hybrid scenarios where both additive and multiplicative systematics are modulating the observed data. Furthermore, even under purely additive or multiplicative systematics, the hybrid algorithm does not perform much worse than the corresponding additive/multiplicative approaches. Thus our hybrid approach becomes an optimal tool in all those scenarios where the additive or multiplicative character of certain systematics is not known a priori. However, in cases where contamination by systematics is minor, and consequently their modulation of the observed number density of galaxies is very mild, the performance of the additive OLS implementation approaches that of the hybrid one: this makes sense provided that, in this limit, all higher order terms in Eq.~\ref{eq:smodel2} involving the combined effect of additive and multiplicative systematics become negligible, and the leading, remaining term is the linear one ($\vzeta\cdot\delta \vM$), since the subtle variance modulation (via the $\bar{n}_g\delta_g(\vbeta\cdot\delta\vM$ term) impacts at next to leading order.

This limiting exercise is interesting since it points to an intrinsic limitation of all the methodologies analysed in this work: there exists a threshold below which none of the methods we are implementing, which are based upon the OLS, is able to correct for systematics. This is hinted in Fig.~\ref{fig:sevtemps_0p02}, where cases in which the level of residuals in the corrected maps do not lie far below the un-corrected cases. In all scenarios depicted in this plot, the residuals on the large angular scales ($\ell\sim 10$) are above or at the level of 5--10\%, regardless which particular methodology is used. 

This is further explored in Fig.\ref{fig:allskytsts}, where we consider a template set that is built upon log-normal simulations of the angular power spectra of the real template set built from J-PLUS DR3 data. These simulated systematics template set shares the angular clustering properties of the real templates, but extend to the entire celestial sphere. This allows us to compare the performance of the three methods under analysis under full sky coverage and under different levels of impact from systematics (from milder to stronger in left to right panels in the plot). In this configuration we find again that the hybrid method outperforms the other two when the impact of systematics is above a given threshold that lies around $\epsilon_i,\beta_i\sim 0.01$. For lower values of $\epsilon_i,\beta_i$ none of the methods is able to notice the presence of the systematics modulation. In all cases, the largest angular scales ($\ell\lesssim 10$) show residuals in the angular power spectra at the level of 2--20\%, below the residual levels found under the J-PLUS DR3 footprint: the amount of available area thus impacts the precision to which we can correct for systematics. Interestingly, one can see that the level of residuals/uncertainties at $\ell=5$ is lower for $\epsilon_{5-11}=\beta_{5-11}=0.02$ (when the OLS actually {\it detects} the systematics impact) than for $\epsilon_{5-11}=\beta_{5-11}=0.005$ (when the OLS fails to detect any systematics-induced modulation). This result highlights the difficulty at correcting the observed angular power on the largest angular scales, which are sensitive to cosmological parameters like, e.g., the inflation-motivated local non-Gaussianity parameter $f_{\rm NL}$. In our simulations the amplitude of the relative residuals seem to decrease with the multipole $\ell$ linearly, i.e., $\propto \ell^{-1}$, which constitutes a stronger decline than the naively expected $\ell^{-1/2}\propto ({\rm no.\,deg.\,freedom})^{-1/2}$ scaling. In reality, the actual scaling is a combined effect depending not only on the amount available degrees of freedom (or $m$ multipoles for a fixed $\ell$), but also on the ratio of angular power spectra of {\it noise} in our algorithms (i.e., $n_g (\vnh)$) and the systematics templates ($\vM (\vnh)$), which is, a priori, a function of angular scale as well. 

We must stress that these results have been obtained under a set of simplifying assumptions, namely (1) uncertainties in the average number of galaxies have been neglected (although they typically re-scale the observed $C_{\ell}$'s with a constant, offset factor), (2) systematics modulate the observed number density of galaxies either in an additive or multiplicative way (although there may be more involved forms in which observing conditions could impact measurements), (3) the re-scaling of the template set $\vM (\vnh)$ versus $n_g(\vnh)$ introduced in Sect.~\ref{sec:method} {\it linearizes} the systematics impact on measurements in a way that makes the model given in Eq.~\ref{eq:smodel1} realistic and accurate, and (4) our systematic template set is complete (in the sense that there is no agent that can potentially impact measurements and that is not included in our template set). One should also bear in mind that the more potential systematics templates are included in the analyses, the more likely it is that {\em random} (i.e., non physical) alignments arise between the template set and the observed galaxy density field, giving rise to spurious corrections on the observations that should not be applied. As emphasized in \citet{weave+21}, the identification of potential systematics and their inclusion in our systematics-correction pipelines is a critical step that must be well motivated on the basis of the observing conditions of each survey. 

Our results compare generally worse (in the sense they predict higher residual levels) than those of \citet{weave+21}, also in the particular case of multiplicative systematics observed in that work. This is partially due to the actual choice of systematics templates. In our case, some systematics templates turned out to be unstable when implementing the multiplicative correction of Eq.~\ref{eq:mult1}, and were detected at relatively lower signal-to-noise ratio by the OLS (partially due to the larger $\ell_{\rm max}$ value in their case, and also partially due to the smaller sky coverage under the J-PLUS DR3 footprint, $f_{\rm sky}=0.09$ versus $f_{\rm sky}=1$ in that work). A lower signal-to-noise ratio in the OLS step reflects on larger errors at all multipoles/scales.

We conclude this section stressing the difficulty associated to the deconvolution of observations from systematics, either instrumental or astrophysical. Many of methods in the literature rely on the OLS algorithm, which itself assumes the existence of a spatial/angular template set $\vM(\vnh)$ that may be imperfect and/or incomplete, and yet the outcome of those methods leave residuals that in many cases cannot be neglected.

\section{Conclusions}
\label{sec:conclusions}

In this work we have addressed, for the first time, the problem of distinguishing between additive and multiplicative systematics impacting the observed number density of matter probes in surveys of the LSS in the universe. We have first implemented the OLS algorithm to identify which spatial templates built upon systematics actually impact the observed number density of galaxies, and then we have followed a minimum variance argument to isolate the multiplicative character of each of those templates.  Once corrected for the multiplicative part of the modulation of the observed galaxy density field, we have applied again the OLS method to correct for the remaining, additive part.

Our algorithm starts with a set of spatial templates with arbitrary angular covariance properties that are built upon a corresponding set of potential systematics.  These templates are confronted to the observed galaxy density and re-scaled to assure a smooth scaling and close-to-linear scaling between templates and galaxy density. The standard OLS method is used to select which templates are actually impacting the data: this subset is then fed into our pipeline that separates the multiplicative from the additive components. 

In this work we have used a particular set of templates inferred from the observing conditions in the J-PLUS DR3 data catalogues, and our methodology has been tested in an ensemble of log-normal mocks whose angular power spectrum has been inferred from real J-PLUS DR3 galaxies. The analysis of these mocks has shown that our hybrid method is able to correct for systematics more accurately than standard methods assuming an additive or multiplicative character of systematics, at least in those cases where systematics were biasing the observed number of galaxies in a hybrid manner (i.e., both in an additive and multiplicative way). In other, more simplistic scenarios where systematics impacting the galaxy density field were purely additive (multiplicative), our hybrid method would perform sub-optimally, but not far from the results obtained with methods assuming a correspondingly additive (multiplicative) character of systematics. 

Our analyses have also shown that whenever the impact of systematics on the observed galaxy field is very mild (or the output of the OLS yields weak or no evidence for all templates), then the (simpler) method assuming additive systematics yields the lowest bias and lowest uncertainty in the corrected angular power spectra. It can be shown that (Eq.~\ref{eq:smodel2}), in this limit, the effective modulation of the observed galaxy field indeed simplifies to the additive scenario. 

We have also found regardless of which particular methodology we have used, there exists a threshold below which the OLS cannot detect the impact of any given systematics template, regardless whether it is multiplicative or additive. This reflects in an unavoidable uncertainty in the angular power spectra that typically increases towards large angular scales, with a scaling that depends upon the ratio of the galaxy and systematics template angular power spectra (but typically close to $\ell^{-\gamma}$, with $\gamma\sim 1$ more positive than the $\gamma=1/2$ value expected from arguments on the number of degrees of freedom). In practice, power spectrum multipoles in the range $\ell \in [2,20]$ typically suffer from uncertainties at the 1--20~\% level, which is still below the cosmic variance limit ($\sim 2/(2\ell+1)/f_{\rm sky}\times 100$~\%).

Nonetheless, our results have obtained under the optimistic assumptions of (1) arbitrarily precise knowledge of the angular average number density of galaxies (which acts as an amplitude scaling of the angular power spectrum), (2) effective additive, multiplicative, or hybrid (additive + multiplicative) modulation of the observed galaxy density field by systematics, (3) correct mapping of systematics impacting observations onto effective sky templates, and (4) a complete set of systematics templates that completely capture the impact of the latter on the constructed galaxy density fields.

In our forthcoming work we shall apply all this methodology on the real data from the J-PLUS DR3 survey.

%________________________________________________________________
\section*{acknowledgements}

C.H.-M. acknowledges the hospitality of the Centro de Estudios de Física del Cosmos de Aragón (ceFca), where part of this work took place. C.H.-M. also acknowledges the support of the Spanish Ministry of Science and Innovation projects PID2021-126616NB-I00, PID2022-142142NB-I00, the European Union through the grant ``UNDARK'' of the Widening participation and spreading excellence programme (project number 101159929), and the contribution from the IAC High-Performance Computing support team and hardware facilities. VM thanks CNPq (Brazil) and FAPES (Brazil) for partial financial support. ET acknowledges funding from the HTM (grant TK202), ETAg (grant PRG1006) and the EU Horizon Europe (EXCOSM, grant No. 101159513). P.A.-M. acknowledges the support of the research Project PID2023-149420NB-I00 funded by MICIU/AEI/10.13039/501100011033 and by ERDF/EU, of the MICIU with funding from the European Union NextGenerationEU and Generalitat Valenciana in the call Programa de Planes Complementarios de I+D+i (PRTR 2022) Project (VAL-JPAS) reference ASFAE/2022/025, and by the project of excellence PROMETEO CIPROM/2023/21 of the Conselleria de Educación, Universidades y Empleo (Generalitat Valenciana). AE acknowledges the financial support from the Spanish Ministry of Science and Innovation and the European Union - NextGenerationEU through the Recovery and Resilience Facility project ICTS-MRR-2021-03-CEFCA. Based on observations made with the JAST80 telescope at the Observatorio Astrofísico de Javalambre (OAJ), in Teruel, owned, managed, and operated by the Centro de Estudios de Física del Cosmos de Aragón. We acknowledge the OAJ Data Processing and Archiving Unit Department\citep[DPAD,][]{davidCH12} for reducing and calibrating the OAJ data used in this work.
Funding for the J-PLUS Project has been provided by the Governments of Spain and Aragón through the {\it Fondo de Inversiones de Teruel}; the Aragón Government through the Research Groups E96, E103, E16\_17R, E16\_20R, and E16\_23R; the Spanish Ministry of Science, Innovation and Universities (MCIN/AEI/10.13039/501100011033 y FEDER, Una manera de hacer Europa) with grants PID2021-124918NB-C41, PID2021-124918NB-C42, PID2021-124918NA-C43, PID2021-124918NB-C44, PGC2018-097585-B-C21 and PGC2018-097585-B-C22; the Spanish Ministry of Economy and Competitiveness (MINECO) under AYA2015-66211-C2-1-P, AYA2015-66211-C2-2, AYA2012-30789, and ICTS-2009-14; and European FEDER funding (FCDD10-4E-867, FCDD13-4E-2685). The Brazilian agencies
FINEP, FAPESP, and the National Observatory of Brazil have also contributed to this project. Some of the results in this paper have been derived using the HEALPix \citep{healpix} package.

%%%%%%%%%%%%%%%%%%%%%%%%%%%%%%%%%%%%%%%%%%%%%%%%%%
\section*{Data Availability}

 All original J-PLUS data can be accessed at \url{https://archive.cefca.es/catalogues}. The software implementing the Hybrid approach presented in this work can be accessed via direct request to the first author, and will be eventually accessible in the github site \url{https://github.com/chmATiac}.

\bibliographystyle{mnras}
%\bibliography{references_pw,pip13II,refs_kDolagI,refs_kDolagII,Planck_bib}
\bibliography{biblio}

%%%%%%%%%%%%%%%%% APPENDICES %%%%%%%%%%%%%%%%%%%%%

\appendix

\section{Monte-Carlo Markov Chain (MCMC) variance minimisation in the HYBRID approach}
\label{sec:appB}

Following the approach outlined in Sect.~\ref{sec:method}, and given a set of systematics templates $\delta \vM (\vnh)$, we attempt to find the configuration of the $\beta_i$'s (or vector $\vbeta$) such that minimizes the variance of the map (see Eq.~\ref{eq:corrmap2a}):
\begin{equation}
n_g^{[2]} (\hat{\beta_i},\vnh) = \frac{n_g^{[1]} (\vnh) }{\prod_i^{N_{\rm act}} (1+\hat{\beta}_i \delta M_i (\vnh)) }. 
\label{eq:corrmap2a_bis}
\end{equation}
The expression for $n_g^{[1]}(\vnh)$ is given in Eq.~\ref{eq:corrmap1} in Sect.~\ref{sec:method}. We define an initial $\chi^2$ statistics given by 
\begin{equation}
\chi^2= \frac{ \biggl({\rm Var}\bigl[n_g^{[2]}(\hat{\beta}_i,\vnh)\bigr] 
%- 0.2\times {\rm Var}\bigl[n_g^{[1]}(\vnh)\bigr] 
\biggr)^2 }{\xi^2}.
\label{eq:ini_chisq}
\end{equation}
In this equation, the term $\xi^2$ in the denominator is obtained after producing a set of $N_{n}$ (log-Normal) mocks having the same angular power spectrum as $n_g^{[1]} (\vnh)$. For each of these mocks, which we also denote as $n_g^{[1]}(\vnh)$, we produce a random set of $N_{\beta}$ {\it small} values of the $\hat{\beta}_i$s, and choose those $\hat{\beta}_i$ values yielding the minimum variance of the map $n_g^{[2]} (\hat{\beta_i},\vnh)$ resulting from Eq.~\ref{eq:corrmap2a_bis}. This variance is typically smaller than the variance of the initial log-Normal galaxy mock denoted as $n_g^{[1]} (\vnh)$. This difference of variance estimates, $\delta \sigma^2 = {\rm Var}[n_g^{[2]}]-{\rm Var}[n_g^{[1]}]$, can be computed for each of the $N_n$ galaxy mocks we are generating. We define $\xi^2$ as the square of the average value of $\delta \sigma^2$ throughout the $N_n$ galaxy mocks we have created:
\begin{equation}
\xi^2:=\bigl( \langle \delta \sigma^2\rangle_{N_n} \bigr)^2.
\label{eq:xidef}
\end{equation}
Thus $\xi^2$ reflects the amount by which the variance of $n_g^{[2]} (\hat{\beta_i},\vnh)$ may actually be below that of $n_g^{[1]} (\vnh)$ for purely random configurations of the $\hat{\beta}_i$s. We emphasize that our galaxy mock sample playing the role of $n_g^{[1]}$ has no systematics, and thus in general one expects that the variance of $n_g^{[2]}$ is higher than that of $n_g^{[1]}$. In this way, $\xi^2$ provides an idea of the amount by which the variance of $n_g^{[2]}$ can be below that of $n_g^{[1]}$ due to pure random alignments of the $\beta_i$s. 
%The pre-factor equal to $0.2$ in the numerator of Eq.~\ref{eq:ini_chisq} is somewhat arbitrary, but it has very limited effects when searching for the $\hat{\beta}_i$ configuration minimizing the variance of $n_g^{[2]}(\hat{\beta}_i,\vnh)$ \\

\begin{comment}
We use the {\tt emcee} python package \citep{emcee} to run MCMC chains and find the $\hat{\beta}_i$ configurations that yield minimum variance of $n_g^{[2]}(\hat{\beta}_i,\vnh)$. Once the MCMC is run with the $\chi^2$ definition given in Eq.~\ref{eq:ini_chisq}, we redefine the $\chi^2$ as
\begin{equation}
\chi^2= \frac{ \biggl({\rm Var}\bigl[n_g^{[2]}(\hat{\beta}_i,\vnh)\bigr] - \min{\bigl( {\rm Var}\bigl[n_g^{[2]}({\hat\beta}_i,\vnh)\bigr]\bigr)} \biggr)^2 }{\xi^2},
\label{eq:ini_chisq2}
\end{equation}
where the ${\rm min}()$ function refers to the minimum variance configuration obtained throughout the MCMC samples. 
From this definition, the probability of any given $\beta_i$ configuration can be re-computed as $\propto \exp{-\chi^2}$. This redefinition however does not change the configuration of the $\hat{\beta}_i$s that yield the minimum variance in Eq.~\ref{eq:corrmap2a_bis}.
\end{comment}

\end{document}